\documentclass[acmlarge,nonacm]{acmart}
\usepackage[utf8]{inputenc}

\usepackage{booktabs} 

\usepackage[ruled]{algorithm2e} 

\acmJournal{TCPS}
 \acmYear{2018}

\setcopyright{acmcopyright}

\acmDOI{0000001.0000001}


\def\mode{0}  

\def\redactionmode{2} 

\def\alttextmode{1} 
\usepackage{adjustbox}              
\usepackage{amsmath}
\usepackage{cleveref}
\usepackage{comment}  				

\usepackage{enumitem}				
\usepackage{graphicx}
	\graphicspath{{./figures/}}		
\usepackage{multirow}				
\usepackage{tabularx} 				
\usepackage{threeparttable}
\usepackage{subfig}
\usepackage{soul} 					
\usepackage{xargs}    				

\usepackage{bchart}

\usepackage{tikz}

\newcommand{\red}{}
\newcommand{\green}{}
\newcommand{\blue}{}

\newcounter{documentmode}
\includecomment{conf}  	
\newcommand{\setdocmode}{%
  \ifcase\number\value{documentmode}
	\renewcommand{\red}[1]{\color{red}##1\color{black}{}}
	\renewcommand{\green}[1]{\color{green}##1\color{black}{}}
	\renewcommand{\blue}[1]{\color{blue}##1\color{black}{}}
  \or
    \usepackage{draftwatermark}
	\SetWatermarkScale{1.35}
	\SetWatermarkAngle{56}
	\SetWatermarkLightness{0.9}
	\SetWatermarkHorCenter{0.9\textwidth}
	\SetWatermarkVerCenter{0.7\textheight}
  \or
	\usepackage{draftwatermark}
	\SetWatermarkText{CONFIDENTIAL}
	\SetWatermarkScale{0.67}
	\SetWatermarkAngle{56}
	\SetWatermarkLightness{0.9}
	\SetWatermarkHorCenter{0.6\textwidth}
	\SetWatermarkVerCenter{0.6\textheight}
  \or
	\usepackage{draftwatermark}
	\SetWatermarkText{Pre-Publication}
	\SetWatermarkScale{2.67}
	\SetWatermarkAngle{56}
	\SetWatermarkLightness{0.9}
	\SetWatermarkHorCenter{0.6\textwidth}
	\SetWatermarkVerCenter{0.6\textheight}
  \else
    \excludecomment{conf}
  \fi
}

\if\mode0
\usepackage[prependcaption, textsize=footnotesize]{todonotes}
\else
\usepackage[disable, prependcaption, textsize=footnotesize]{todonotes}

\fi

\setdocmode

\newcounter{redmode}
\includecomment{2Detailed}			
\includecomment{Uncritical}			
\includecomment{2SaveSpace}			
\includecomment{IfMustDelete}			
\newcommand{\redact}{%
  \ifcase\number\value{redmode}
    \includecomment{2Detailed}
    \includecomment{Uncritical}
    \includecomment{2SaveSpace}
    \includecomment{IfMustDelete}
  \or
    \excludecomment{2Detailed}
    \includecomment{Uncritical}
    \includecomment{2SaveSpace}
    \includecomment{IfMustDelete}
  \or
    \excludecomment{2Detailed}
    \excludecomment{Uncritical}
    \includecomment{2SaveSpace}
    \includecomment{IfMustDelete}
  \or
    \excludecomment{2Detailed}
    \excludecomment{Uncritical}
    \excludecomment{2SaveSpace}
    \includecomment{IfMustDelete}
  \or
    \excludecomment{2Detailed}
    \excludecomment{Uncritical}
    \excludecomment{2SaveSpace}
    \excludecomment{IfMustDelete}
  \fi
}
\redact

\newcounter{alttxtmode}
\includecomment{TextA}			
\includecomment{TextB}			
\newcommand{\alttext}{%
  \ifcase\number\value{alttxtmode}
    \includecomment{TextA}
    \includecomment{TextB}
  \or
    \excludecomment{TextA}
    \includecomment{TextB}
  \or
    \includecomment{TextA}
    \excludecomment{TextB}
  \or
    \excludecomment{TextA}
    \excludecomment{TextB}
  \fi
}
\alttext

\usepackage[acronym]{glossaries} 

\newacronym{am}{AM}{Amplitude Modulation}
\newacronym{asd}{ASD}{Autism Spectrum Disorder}
\newacronym{at}{AT}{Adaptive Tracking}
\newacronym{bw}{BW}{Baseline Wanderer}
\newacronym{bpm}{BPM}{Breath per Minute}
\newacronym{bvp}{BVP}{Blood Volume Pulse}
\newacronym{cemd}{CEMD}{Complex Empirical Mode Decomposition}
\newacronym{co}{CO}{Count Origin}
\newacronym{cosf}{TDCO}{Time Domain - Count Origin}
\newacronym{costf}{COSTF}{Count Origin - Smart and Time Fusion}
\newacronym{cwt}{CWT}{Continuous Wavelet Transform}
\newacronym{der}{DER}{Detection Error Rate}
\newacronym{df}{DF}{Dominant Frequency}
\newacronym{dfsf}{SFFDP}{Smart Fusion of Frequency Domain Peak}
\newacronym{dwt}{DWT}{Discrete Wavelet Transform}
\newacronym{ecg}{ECG}{Electrocardiography}
\newacronym{skin_eda}{EDA}{Electrodermal activity} 
\newacronym{eeg}{EEG}{Electroencephalogram}
\newacronym{emd}{EMD}{Empirical Mode Decomposition}
\newacronym{emg}{EMG}{Electromyography}
\newacronym{eog}{EOG}{Electrooculogram}
\newacronym{ews}{EWS}{Early Warning Score}
\newacronym{fft}{FFT}{Fast Fourier Transformation}
\newacronym{fm}{FM} {Frequency Modulation}
\newacronym{gsr}{GSR}{Galvanic Skin Response}
\newacronym{hf}{HF}{High Frequency}
\newacronym{hr}{HR}{Heart Rate}
\newacronym{hrv}{HRV}{Heart rate variability}
\newacronym{ibi}{IBI}{Inter beat interval}
\newacronym{ica}{ICA}{Independent Component Analysis}
\newacronym{iir}{IIR}{Infinite Impulse Response}
\newacronym{imf}{IMF}{Intrinsic Mode Function}
\newacronym{knn}{KNN}{k-nearest neighbour}
\newacronym{lda}{LDA}{Linear Discriminant Analysis}
\newacronym{led}{LED}{light emitting diode}
\newacronym{lms}{LMS}{Least Mean Square}
\newacronym{lf}{LF}{low frequency}
\newacronym{ma}{MA}{Movement Artifact}
\newacronym{mad}{MAD}{Mean Absolute Deviation} 
\newacronym{mar}{MAR}{Motion Artifact Removal} 
\newacronym{mimic}{MIMIC}{Multi-parameter Intelligent Monitoring for Intensive Care}
\newacronym{nlms}{NLMS}{Normalized Least Mean Square}
\newacronym{n2n}{N2N}{normal-to-normal}
\newacronym{pd}{PD}{Peak Detection}
\newacronym{pdsf}{TDPD}{Time Domain Peak Detection}
\newacronym{ppe}{PPE}{Peak to Peak Error}
\newacronym{ppg}{PPG}{Photoplethysmogram}
\newacronym{ptt}{PTT}{Pulse transit time}
\newacronym{rls}{RLS}{ Recursive Least Square}
\newacronym{rms}{RMS}{ Root Mean Square}
\newacronym{rmse}{RMSE}{Root Mean Square Error}
\newacronym{rr}{RR}{Respiratory Rate}
\newacronym{rrsv}{RRSV}{Repeated Random Sub-Sampling Validation}
\newacronym{sam}{SAM}{Self-Assessment Manikin}
\newacronym{saews}{SA-EWS}{Self-Aware Early Warning Score}
\newacronym{sart}{SART}{sustained attention to response test}
\newacronym{sfu}{SFU}{Smart Fusion}
\newacronym{scl}{SCL}{Skin Conductance Level}
\newacronym{scr}{SCR}{Skin Conductance Response}
\newacronym{sigvm}{SigVM}{Signal Vector Magnitude}
\newacronym{skt}{SKT}{Skin Temperature}
\newacronym{std}{STD}{Standard Deviation }
\newacronym{svd}{SVD}{Singular Value Decomposition }
\newacronym{tfu}{TFU}{Temporal Fusion } 
\newacronym{vfcdm}{VFCDM} {Variable Frequency Complex Demodulation}
\newacronym{whs}{WHS}{Wearable Health-care Systems}

\newacronym{ai}{AI}{Artificial Intelligence}
\newacronym{ann}{ANN}{Artificial Neural Network}
\newacronym{bpn}{BPN}{Back-Propagation Neural Network}
\newacronym{bpnn}{BPNN}{back-propagation neural network}
\newacronym{cart}{CART}{Classification And Regression Tree}
\newacronym{cnn}{CNN}{Convolutional Neural Network}
\newacronym{flop}{FLOP}{Floating Point Operation}
\newacronym{icnn}{ICNN}{Iterative Convolutional Neural Network}
\newacronym{ldf}{LDF}{Linear Discriminant Function}
\newacronym{mcs}{MCS}{Multiple Classifier System}
\newacronym{ml}{ML}{Machine Learning}
\newacronym{nn}{NN}{Neural Network}
\newacronym{nb}{NB}{Naive Bayesian}
\newacronym{rsvm}{RSVM}{Reputation-driven Support Vector Machine }
\newacronym{svm}{SVM}{Support Vector Machine}
\newacronym{ucnn}{$\mu$CNN}{Micro CNN}

\newacronym{brs}{BRS}{Bipolar Resistive Switch-based logic}
\newacronym{cnf}{CNF}{Conjunctive Normal Form}
\newacronym{crs}{CRS}{Complementary Resistive Switch-based logic}
\newacronym{dnf}{DNF}{Disjunctive Normal Form}
\newacronym{fpm}{FPM}{Forward Polarized Memristor}
\newacronym{hfo}{$HfO_x$}{Hafnium Oxide}
\newacronym{hrs}{HRS}{High Resistance State}
\newacronym{imc}{IMC}{In-Memory Computation}
\newacronym{imply}{IMPLY}{Material Implication}
\newacronym{imp}{IMP}{In-Memory Processing}
\newacronym{lim}{LIM}{Logic in Memory}
\newacronym{lrs}{LRS}{Low Resistance State}
\newacronym{ltg}{LTG}{Logic Threshold Gate}
\newacronym{magic}{MAGIC}{Memristor-Aided Logic}
\newacronym{mecoins}{Me-Coin}{Memristor-based Computation In-memory}
\newacronym{pcm}{PCM}{Phase Change Memory}
\newacronym{pim}{PIM}{Processing in Memory}
\newacronym{reram}{ReRAM}{Resistive Random Access Memory}
\newacronym{rpm}{RPM}{Reversely Polarized Memristor}
\newacronym{sdc}{SDC}{Self Directed Channel}
\newacronym{stt}{STT}{Spin Transfer Torque}
\newacronym{tao}{$TaO_x$}{Tantalum Oxide}
\newacronym{tio}{$TiO_2$}{Titanium dioxide}
\newacronym{vteam}{VTEAM}{Voltage-controlled ThrEshold Adaptive Memristor}

\newacronym{aco}{ACO}{Autonomous Cooperating Object}
\newacronym{afdd}{AFDD}{Automated Fault Detection and Diagnostic}
\newacronym{ca}{CA}{Continuous Average}
\newacronym{cah}{CAH}{Context-Aware Health Monitoring}
\newacronym{cam}{CCAM}{Confidence-based Context-Aware condition Monitoring}
\newacronym{csa}{CSA}{Computational Self-Awareness}
\newacronym[plural=DABs,longplural={Discrete Average Blocks}]{dab}{DAB}{Discrete Average Block}
\newacronym[plural=KPNs,longplural={Kahn Process Networks}]{kpn}{KPN}{Kahn Process Networks} 
\newacronym{mape-k}{MAPE-K}{Monitor-Analyze-Plan-Execute over a shared Knowledge}
\newacronym[plural=MoCs,longplural={Models of Computation}]{moc}{MoC}{Model of Computation}
\newacronym{oda}{ODA}{Observe-Decide-Act}
\newacronym{pca}{PCA}{Principal Component Analysis}
\newacronym{rosa}{RoSA}{Research on Self-Awareness}
\newacronym{sa}{SA}{Self-Aware}
\newacronym{saness}{SA}{Self-Awareness}
\newacronym{sahm}{SAHM}{Self-Aware Health Monitoring}
\newacronym{samba}{SAMBA}{Self-Aware health Monitoring and Bio-inspired coordination for distributed Automation systems}
\newacronym{selphys}{SelPhyS}{Self-aware cyber-Physical System}
\newacronym{sh}{SH}{State Handler}
\newacronym{som}{SOM}{Self-Organizing Map}

\newacronym{c-s}{CAS}{Compare-and-Swap}
\newacronym{cps}{CPS}{Cyber-Physical System}
\newacronym{cpps}{CPPS}{Cyber-Physical Production System}
\newacronym{dsr}{DSR}{Down-Sampling Rate}
\newacronym{dum}{DuM}{Device under Monitoring}
\newacronym{es}{ES}{Embedded System}
\newacronym{LD}{LD}{low-discrepancy}
\newacronym{mes}{MES}{Manufacturing Execution System}
\newacronym{sc}{SC}{Stochastic Computing}
\newacronym{sos}{SoS}{System of Systems}
\newacronym{suo}{SuO}{System under Observation}

\newacronym{abi}{ABI}{Application Binary Interface}
\newacronym{adc}{ADC}{Analog-to-Digital Converter}
\newacronym{aes}{AES}{Advanced Encryption Standard}
\newacronym{alu}{ALU}{Arithmetic Logic Unit}
\newacronym{api}{API}{Application Programming Interface}
\newacronym{asic}{ASIC}{Application Specific Integrated Circuit}
\newacronym{asoc}{ASOC}{Autonomic System-on-Chip platform}
\newacronym{axi}{AXI}{Advanced eXtensible Interface Bus}
\newacronym{bram}{BRAM}{Block Random Access Memory}
\newacronym{cdt}{CDT}{C/C++ Development Tooling}
\newacronym{clb}{CLB}{Configuarable Logic Block}
\newacronym{cmos}{CMOS}{Complementary Metal-Oxide Semiconductor}
\newacronym{cp}{CP}{Clock Pulse}
\newacronym{cpi}{CPI}{Cycles Per Instruction}
\newacronym{cpu}{CPU}{Central Processing Unit} 
\newacronym{cpsoc}{CPSoC}{Cyber-Physical System-on-Chip}
\newacronym{cu}{CU}{Compute Unit}
\newacronym{cuda}{CUDA}{Compute Unified Device Architecture}
\newacronym{dac}{DAC}{Digital to Analog Converter}
\newacronym{ddr3}{DDR3}{Double Data Rate}
\newacronym{dff}{DFF}{Data Flip-Flop}
\newacronym{dll}{DLL}{Delay Locked Loop}
\newacronym{dmr}{DMR}{Dual Modular Redundancy}
\newacronym{dram}{DRAM}{Dynamic Random Access Memory}
\newacronym{dsd}{DSD}{Digital Synchronous Detection}
\newacronym{dsp}{DSP}{Digital Signal Processor}
\newacronym{dt}{DigiTime}{}
\newacronym{dvfs}{DVFS}{Dynamic Voltage and Frequency Scaling}
\newacronym{eda}{EDA}{Electronic Design Automation}
\newacronym{fdc}{FDC}{Frequency-to-Digital Converter}
\newacronym{fifo}{FIFO}{First In First Out}
\newacronym{fpga}{FPGA}{Field Programmable Gate Array}
\newacronym{gds}{GDS}{Global Data Share}
\newacronym{gnulgpl}{GNU LGPL}{GNU Lesser General Public Licence} 
\newacronym{gpgpu}{GPGPU}{General Purpose Graphics Processing Unit}
\newacronym{gpr}{GPR}{General Purpose Register}
\newacronym{gpu}{GPU}{Graphics Processing Unit}
\newacronym{gro}{GRO}{Gated Ring Oscillator}
\newacronym{io}{IO}{Input-Output}
\newacronym{hamsoc}{HAMSoC}{Hierarchical Agent Monitoring System-on-Chip}
\newacronym{hdl}{HDL}{Hardware Description Language}
\newacronym{hmp}{HMP}{Heterogeneous Multi-Processor}
\newacronym{ic}{IC}{Integrated Circuit}
\newacronym{icap}{ICAP}{Internal Configuration Access Port}
\newacronym[longplural={Intellectual Properties}]{ip}{IP}{Intellectual Property}
\newacronym{isa}{ISA}{Instruction Set Architecture}
\newacronym{lds}{LDS}{Local Data Share}
\newacronym{lru}{LRU}{Least Recently Used}
\newacronym{lsb}{LSB}{Least-Significant Bit}
\newacronym{lsu}{LSU}{Load Store Unit}
\newacronym{lut}{LUT}{Look Up Table}
\newacronym{mash}{MASH}{Multi-Stage Noise-Shaping}
\newacronym{mems}{MEMS}{Micro-Electro-Mechanical Systems}
\newacronym{miaow}{MIAOW}{Many-core Integrated Accelerator Of deepwater/Wisconsin}
\newacronym{mosfet}{MOSFET}{Metal Oxide Semiconductor Field Effect Transistor}
\newacronym{mpsoc}{MPSoC}{Multi-Processor System-on-Chip}
\newacronym{mshr}{MSHR}{Miss Status Holding/Handling Register}
\newacronym{noc}{NoC}{Network-on-Chip}
\newacronym{opencl}{OpenCL}{Open Computing Language}
\newacronym{ocn}{OCN}{On-Chip Network}
\newacronym{pcb}{PCB}{Printed Circuit Board}
\newacronym{pcie}{PCIe}{Peripheral Component Interconnect Express}
\newacronym{pl}{PL}{Programmable Logic}
\newacronym{pli}{PLI}{Verilog Programming Language Interface}
\newacronym{pll}{PLL}{Phase-Locked Loop}
\newacronym{ps}{PS}{Processing System}
\newacronym{pv}{PV}{Process Variation}
\newacronym{qoe}{QoE}{Quality of Experience}
\newacronym{qos}{QoS}{Quality of Service}
\newacronym{ram}{RAM}{Random Access Memory} 
\newacronym{risc}{RISC}{Reduced Instruction Set Computer}
\newacronym{riscv}{RISC-V}{Reduced Instruction Set Computing - V}
\newacronym{rtl}{RTL}{Register-Transfer Level}
\newacronym{sdk}{SDK}{Software Development Kit}
\newacronym{seec}{SEEC}{SElf-awarE Computing}
\newacronym{sgpr}{SGPR}{Scalar General Purpose Register}
\newacronym{si}{SI}{Southern Island}
\newacronym{simd}{SIMD}{Single Instruction Multiple Data}
\newacronym{simf}{SIMF}{Single Instruction Multiple Floating point}
\newacronym{sm}{SM}{Streaming Multiprocessor}
\newacronym{snr}{SNR}{Signal to Noise Ratio}
\newacronym[plural=SoCs,firstplural=Systems on Chip (SoCs)]{soc}{SoC}{System-on-Chip}
\newacronym{spared}{SPARED}{Self-aware PArtial Reconfiguration architecture for Edge Devices}
\newacronym{spice}{SPICE}{Simulation Program With Integrated Circuit Emphasis}
\newacronym{tad}{TAD}{Time \gls{adc}}
\newacronym[plural=TDCs,longplural={Time-to-Digital Converters}]{tdc}{TDC}{Time-to-Digital Converter}
\newacronym{tq}{TQ}{Time-Quantizer}
\newacronym{uart}{UART}{Universal Asynchronous Receiver/Transmitter}
\newacronym{vcdu}{VCDU}{Voltage Controlled Delay Unit}
\newacronym{vco}{VCO}{Voltage Controlled Oscillator}
\newacronym{vga}{VGA}{Video Graphics Array}
\newacronym{vhdl}{VHDL}{Very High Speed Integrated Circuit Hardware Description Language}
\newacronym{vlsi}{VLSI}{Very Large Scale Integration}
\newacronym{vgpr}{VGPR}{Vector General Purpose Register}
\newacronym{xilffs}{XILFFS}{Generic Fat File System Library}

\newacronym{amd}{AMD}{Advanced Micro Devices}
\newacronym{beol}{BEOL}{Back End Of Line}
\newacronym{cad}{CAD}{Computer-Aided Design}
\newacronym{cas}{CAS}{Circuits and Systems}
\newacronym{dfa}{DFA}{Discriminant Function Analysis} 
\newacronym{eu}{EU}{European Union}
\newacronym{fdd}{FDD}{fault detection and diagnostic}
\newacronym{fefet}{FeFET}{Ferroelectric Field Effect Transistor}
\newacronym{feline}{FeLINe}{FeFET Logic IN mEmory}
\newacronym[plural=FoMs,longplural={Figures of Merit}]{fom}{FoM}{Figure of Merit}
\newacronym{ga}{GA}{Genetic Algorithm}
\newacronym{hipeac}{HiPEAC}{High Performance and Embedded Architecture and Compilation}
\newacronym{hp}{HP}{Hewlett Packard}
\newacronym{hqp}{HQP}{Highly Qualified People}
\newacronym{hvac}{HVAC}{Heating, Ventilation and Air Conditioning}
\newacronym{ibm}{IBM}{International Business Machines corporation}
\newacronym{ict}{ICT}{Institute for Computer Technology}
\newacronym{iot}{IoT}{Internet of Things}
\newacronym{nda}{NDA}{Non-Disclosure Agreement}
\newacronym{nvp}{NVP}{Non-Volatile Processor}
\newacronym{oecd}{OECD}{Organization for Economic Cooperation and Development}
\newacronym{rd}{R\&D}{Research and Development}
\newacronym{rmsdd}{RMSDD}{Root-Mean Square of Successive Differences}
\newacronym{sdnn}{SDNN}{Standard Deviation Normal-to-Normal-Intervals}
\newacronym{soa}{SoA}{State-of-the-Art}
\newacronym{tsmc}{TSMC}{Taiwan Semiconductor Manufacturing Company}
\newacronym{tvlsi}{TVLSI}{Transactions on Very Large Scale Integration}

\newacronym{rhr}{RHR}{Resting Heart Rate}
\newacronym{hrvme}{HRV}{Heart Rate Variability}
\newacronym{spo2}{SpO2}{Oxygen Saturation}
\newacronym{bp}{BP}{Blood Pressure}
\newacronym{sbp}{SBP}{Systolic Blood Pressure}
\newacronym{sbpp}{SBPp}{Peripheral systolic blood pressure}
\newacronym{dbp}{DBP}{Diastolic Blood Pressure}
\newacronym{ssbp}{SSPB}{Steepness in the SBP peaks}
\newacronym{edas}{EDA}{Electrodermal Activity}
\newacronym{gps}{GPS}{Global Positioning System}
\newacronym{rsp}{RSP}{Respiration} 
\newacronym{detr}{DT}{Decision Tree}
\newacronym{rf}{RF}{Random Forest}
\newacronym{cafs}{CAFS}{Cost-Aware Features Selection}

\newacronym{psd}{PSD}{Power Spectral Density}
\newacronym{hht}{HHT}{Hilbert Huang Transform}
\newacronym{vae}{VAE}{Variational Autoencoder}

\newacronym{dtc}{DT}{Decision Tree}
\newacronym{mlp}{MLP}{Multilayer Perceptron}
\newacronym{xgb}{XGBoost}{Extreme Gradient Boosting}
\newacronym{vgg}{VGG}{Visual Geometry Group}
\newacronym{elm}{ELM}{Extreme Learning Machine}
\newacronym{adhd}{ADHD}{Attention Deficit Hyperactivity Disorder}
\newacronym{bci}{BCI}{Brain-Computer Interface}
\newacronym{st}{ST}{Skin Temperature}
\newacronym{dnn}{DNN}{Deep Neural Networks}

\include{acronyms_local}

\begin{document}
\title{Wearable Healthcare Devices for Monitoring Stress and Attention Level in Workplace Environments}

%
%



\author{Peter Traunmueller}
\authornote{Both authors contributed equally to this research.}
\email{}
\orcid{0000-0003-2599-6011}
\author{Anice Jahanjoo}
\orcid{0000-0002-1292-8338}
\authornotemark[1]
\email{anice.jahanjoo@tuwien.ac.at}
\affiliation{%
  \institution{TU Wien}
  \department{Institute of Computer Technology}
  \streetaddress{Gusshausstrasse 27-29}
  \city{Vienna}
  \country{Austria}
  \postcode{1040}
}



\author{Soheil Khooyooz}
\orcid{}
\affiliation{
  \institution{Heidelberg University}
  \department{Institute of Computer Engineering (ZITI)}
  \streetaddress{Im Neunenheimer Feld 368}
  \city{Heidelberg}
  \postcode{69120}
  \country{Germany}
  }
\email{soheil.khooyooz@ziti.uni-heidelberg.de}

\author{Amin Aminifar}
\orcid{}
\affiliation{
  \institution{Heidelberg University}
  \department{Institute of Computer Engineering (ZITI)}
  \streetaddress{Im Neunenheimer Feld 368}
  \city{Heidelberg}
  \postcode{69120}
  \country{Germany}
  }
\email{amin.aminifar@ziti.uni-heidelberg.de}

\author{Nima TaheriNejad}
\orcid{0000-0002-1295-0332}
\affiliation{
  \institution{TU Wien \& Heidelberg University}
  \department{Institute of Computer Technology}
  \streetaddress{Gusshausstrasse 27-29}
  \city{Vienna}
  \postcode{1040}
  \country{Austria \& Germany}
  }
\email{nima.taherinejad@tuwien.ac.at \& nima.taherinejad@ziti.uni-heidelberg.de}

\renewcommand\shortauthors{Traunmueller et al}


\begin{abstract}
Wearable devices have revolutionized healthcare monitoring, allowing us to track physiological conditions without disrupting daily routines. Whereas monitoring physical health and physical activities have been widely studied, their application and impact on mental health are significantly understudied. This work reviews the state-of-the-art, focusing on stress and concentration levels. These two can play an important role in workplace humanization. For instance, they can guide breaks in high-pressure workplaces, indicating when and how long to take. Those are important to avoid overwork and burn-out, harming employees and employers. To this end, it is necessary to study which sensors can accurately determine stress and attention levels, considering that they should not interfere with their activities and be comfortable to wear. From the software point of view, it is helpful to know the capabilities and performance of various algorithms, especially for uncontrolled workplace environments. This work aims to research, review, and compare commercially available non-intrusive measurement devices, which can be worn during the day and possibly integrated with healthcare systems for stress and concentration assessment. We analyze the performance of various algorithms used for stress and concentration level assessment and discuss future paths for reliable detection of these two parameters.
\end{abstract}


\maketitle
\glsresetall

\section{INTRODUCTION}
\label{Sec:Introduction}

Many breakthroughs in electronics, information, and communication technology,  particularly \gls{whs}, have provided us with unprecedented access to physiological information. \glspl{whs} are used in many applications, such as well-being and sports activities~\cite{Perego2017, Yin2018, Perego2021}, physical health~\cite{gotzinger2016confidence,  anzanpour:2017a,  Pollreisz2020_LASCAS, Taherinejad2020healthwear, Freismuth2022}, and mental health~\cite{Taherinejad2016emotions, Taherinejad2016_CompObs, Pollreisz2017,  Freismuth2022,jahanjoo2024high}. These systems promise to revolutionize various facets of our society by generating an exceptional understanding of humans and their bodies.
There are many different devices for monitoring and collecting our biosignals ~\cite{garmintactix_spec,e4_specifications_website,muse2_specifications_website}. Therefore, it is important to know which one(s) to use for a given task. Many devices that produce the highest data quality are uncomfortable to wear on a daily basis, intervene with daily activities, or severely invade the users' privacy ~\cite{Taherinejad2020LSCNewsletter, anzanpour:2017a, nonintrusive_vital_sign}.  Others that may be easy to use do not necessarily provide the information, accuracy, or reliability desired or expected from them.

To understand the potentials of \glspl{whs} for a specific purpose or select a \gls{whs} for a particular purpose, sensors (hardware) and algorithms (software) of such systems should be studied. The sensors and algorithms, however, can only be studied in their context and intended applications. Given that studying the entire range of various applications, usage context, and monitoring parameters in a single study is not realistic, we narrow down our focus. Here, we consider wearable healthcare in the context of work environments, primarily that of rapidly growing digital work environments. In such an environment, two physiological parameters stand out. One is the concentration that is connected to efficiency and even the quality of the work. Both employers and employees may be interested in monitoring concentration to improve productivity. However, this may lead to excessive stress, negatively affect employees' health, and consequently degrade productivity, the same factor that was the improvement target. There may be many other factors that can lead to excessive stress in the work environment and lead to unintended consequences for both employees and employers. Therefore, another important parameter that we consider in this paper is stress level assessment. 

Choosing and combining the correct physiological variables plays a significant role in measuring mental workload, which often correlates with stress and attention levels. Having the necessary biomedical sensors for respective physiological parameters is not enough, as interactions of subjects with the environment often limit their accuracy. Therefore, additional methods and sensors for countering such negative effects are needed, e.g., acceleration measurements for removing movement artifacts. Implementing complex algorithms countering the loss of precision may seem intuitive at first; however, there is a compromise to consider since they require processing resources (and thus may slow down the system) and decrease battery life ~\cite{combining_and_comparing, wearable_medical_devices,Shakibhamedan2024EASE}.

Another compromise to consider is that of desired accuracy and the cost of having specialized sensors associated with each of the physiological parameters one desires to monitor. This associated cost is not only monetary but also includes considerations of mechanical aspects such as adding more sensors (space, weight, looks, ease of use) and electrical aspects (additional power consumption). Consequently, using sensors from which multiple vital signs can be (indirectly) extracted is a common technique. 
This means that using just one sensor, multiple physiological parameters are measured. However, such methods may suffer from lower accuracy or be more prone to interference. Although, sometimes, the extracted signs can be jointly considered to achieve higher accuracy~\cite{Goetzinger2019ews}. Most commercial devices already feature a set of sensors, which makes many combinations possible, although finding the best ones is often very difficult ~\cite{reliable_respiratory_rate, combining_and_comparing}. Moreover, smart techniques for data collection may be used to decrease the power consumption of data collection from each sensor~\cite{Taherinejad2017, Hadizadeh2020, Hadizadeh2021}, allowing for more sensors from the power budget (battery capacity) point of view.

Given that in the work environment, wearable devices should be non-intrusive (not to interrupt routine work procedures). As comfortable as possible, we narrowed down the scope of this survey to wrist-worn devices (smartwatches), smart rings, and headbands. We note that there are other devices that could theoretically fit these needs. However, they are either in a maturing phase and not very common (no or too few options available commercially) or not as easy to use, change, or charge in the work environment (e.g., chest straps). 
Most modern wearable healthcare devices feature a specific set of sensors, explained by the limitations mentioned above. In wrist-worn devices, those are \gls{ppg} sensors and accelerometers. \gls{ppg} sensors measure reflected light from a light source on the skin.
Different wavelengths are often used simultaneously to extract multiple physiological parameters and/or more reliable readings. Only a few commercial wearable health monitoring devices 
embed additional sensors such as temperature or skin conductance \cite{wearable_health_devices}.  
For head-worn devices, \gls{eeg} 
is the most prominent sensor used to measure and record brain activity. \gls{eeg} sensors measure differences (voltage differences) at the scalp produced by electrical currents generated via neural activities in the brain. Newer devices are equipped with accelerometers that enable researchers to calculate movement and balance during health monitoring ~\cite{senzeband_specifications_website, muse2_specifications_website,museS_specifications_website}. \gls{ppg} sensors are only implemented in several of them~\cite{wearable_health_devices, wearable_medical_devices}.

Due to the necessary certifications of devices used outside laboratory environments, only commercially available products can be used for work environments. All modern wearable devices offer wireless data transmission to other devices, such as mobile phones and computers, or can directly upload data onto a server. Extracting raw data from these devices is not always possible, and the manufacturer sometimes limits it. Unofficial extraction tools may be an option in the short term but may be rendered unusable at any time due to a software update on the device. Extensive studies have been done concerning the validity of measured data from those devices. If the desired metrics can be reliably obtained through uncomplicated algorithms, their usability increases~\cite{consumer_grade_EEG, detection_and_removal}.

The rest of this paper is organized as follows: In \Cref{sec_wearable_devices}, we present currently available commercial devices, their sensors, and key features. We review various applications in \Cref{applications} with a comprehensive investigation of stress and attention detection, which are the primary subjects of this survey. We discuss our findings in \Cref{sec_discussion} before concluding the paper in \Cref{sec_conclusion}.


\section{Wearable devices} 
\label{sec_wearable_devices}


Even though many different devices are available, most of them focus on just a few wearing positions. Due to physiological constraints, \gls{eeg} measuring devices are head-worn. \gls{ppg} sensors are usually wrist-worn due to their almost independent measurement position in combination with the uncomplicated and unobtrusive integration into smartwatches. This has been shown to decrease the quality of data but is by far the most popular choice for commercial manufacturers ~\cite{HRV_and_stress, a_review_on, consumer_grade_EEG}.

\gls{eeg} combined with other physiological sensor data has proven to be a good match (89\% accuracy) for distinguishing between high and low workloads. The option of combining \gls{eeg} with eye data would have slightly higher accuracy (91\% accuracy) \cite{combining_and_comparing} but can not be used in a workplace environment due to privacy concerns. 

In \Cref{sensor_to_application}, we present a schematic representation of the sensor distribution across categories of common devices. Each category is denoted by a distinct color, highlighting the specific sensors employed. Notably, there is overlap in sensor usage among several devices, indicating shared functionality and potential synergy in data collection.

In light of these constraints and popular choices, wearable devices hold tremendous promise in healthcare systems. Their capacity for continuous monitoring of vital signs and real-time data collection makes them a focal point of interest for enhancing patient care. By exploring their integration in healthcare, we aim to explore various wearable devices and their potential for enhancing healthcare efficiency by utilizing their data.

{\begin{figure}[hbp]
\centerline{\includegraphics[width=0.88\columnwidth]{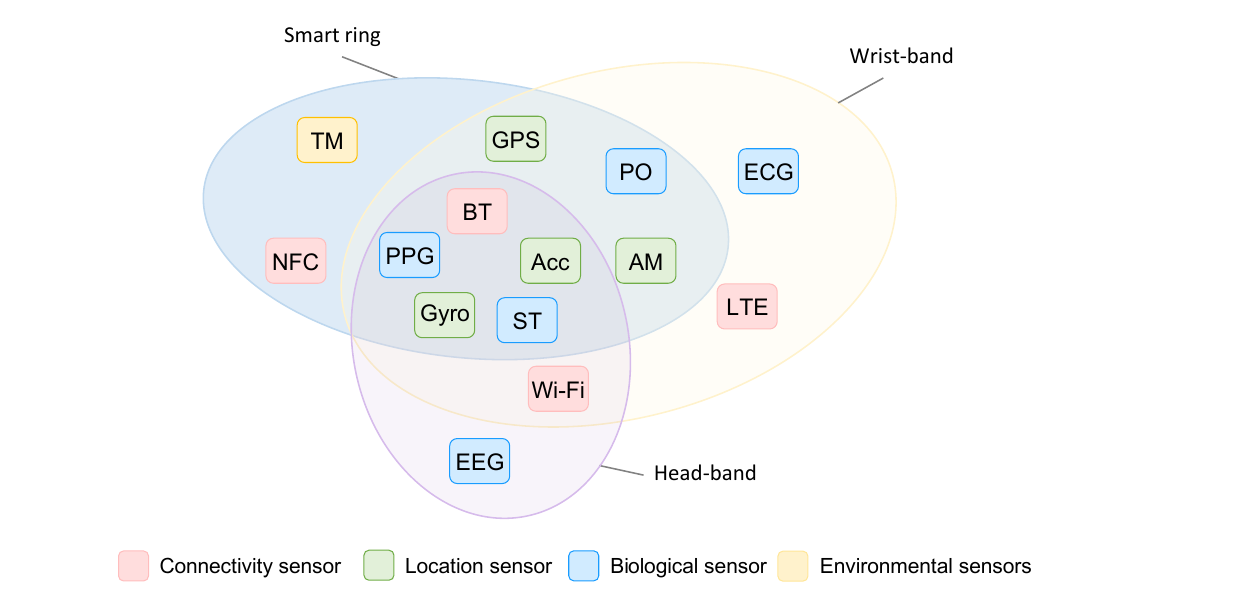}}
\caption{Embedded sensors in wearable devices (PO: Pulse Oximeter, TM: Thermometer, BT: Bluetooth, ST: Skin Temperature, AM: Altimeter, Acc: Accelerometer, Gyro: Gyroscope, GPS: Global Positioning System)
 }
\label{sensor_to_application}
\end{figure}}
\subsection{Head-worn devices}

Head-worn devices are mainly used for \gls{eeg} measurements. Modern head-worn devices include acceleration sensors to remove motion artifacts and sometimes additional sensors like \gls{ppg} or temperature sensor ~\cite{museS_specifications_website, senzeband_specifications_website,muse2_specifications_website}. For a clean \gls{eeg} recording, the signals need to be filtered, and manual artifact removal is usually performed. The filtered signal can be transformed from the time domain into the frequency domain via \gls{fft}, which is essential for categorization, where certain frequency bands have been traditionally defined and used to classify the signal \cite{10052897}. As shown in \Cref{table:eeg_spectrums}, those spectrums directly correlate to stress and attention levels and are therefore a key indicator ~\cite{gamma_alpha_delta, DETECTION_OF_DRIVER, novel_approach_for}. 
In the rest of this subsection, we briefly review major (commercially available) devices, their specifications, capabilities, and sensors.  \Cref{table:head_worn_table} shows a summary of modern head-worn \gls{eeg} devices.

\begin{table*}[t]
\caption{\gls{eeg} frequency ranges~\cite{novel_approach_for} }
\centering
\begin{tabular}{ccl}
\toprule
      \multicolumn{1}{c}{Signal}  & \multicolumn{1}{c}{Frequency {[}Hz{]}} & \multicolumn{1}{c}{Activity}     \\ \midrule 
    $\delta$ Delta & \textless{}4       & Increases during difficult conditions \\
    $\theta$ Theta & 4-8                & Increases during stress               \\
    $\alpha$ Alpha & 8-12               & Decreases during stress           \\
    $\beta$ Beta   & 13-31              & Varies according to task difficulty   \\
    $\gamma$ Gamma & \textgreater{}31   & Increases during meditation \\ 
\bottomrule 
\end{tabular}
\label{table:eeg_spectrums}
\end{table*}

\subsubsection{Neeuro SenzeBand}
    The SenzeBand 2, released in 2021, offers seven \gls{eeg} channels. Two out of the seven sensors are located on the forehead, and two are temporal lobe sensors. It can be used on dry skin, providing a battery lifespan of up to 6 hours. 
    Additionally, a \gls{ppg} and 9-axis motion sensor is built into the unit. There is a dedicated \gls{sdk} for raw data extraction and analysis provided by the manufacturer~\cite{senzeband_specifications_website, consumer_grade_EEG}. 

    \subsubsection{Emotiv EPOC X}

Due to the popularity of their predecessor product EPOC+, Emotiv offers the EPOC X  a 14-channel \gls{eeg} headset. The devices from Emotiv have the highest proportion of publications and a significant market share. By using wet electrodes, the device offers a better signal quality than dry electrodes; however, this makes the headband more complicated to use in daily and work environments. Users have expressed problems with wearing the headset over a longer period since the device offers no adjustability for different head sizes. The internal 9-axis motion sensor and \gls{eeg} data can be extracted via a paid toolkit from Emotiv or unofficial extraction tools \cite{epoc_specifications_website, consumer_grade_EEG}. 

\subsubsection{OpenBCI Cyton EEG Headband}
With each element being individually sold, the OpenBCI headband offers an open-source alternative to other manufacturers. There are four channels of \gls{eeg}, \gls{ecg}, \gls{emg}, which is the recording of electric signals generated during muscle movements~\cite{xiong2021deep} and a 3-axis accelerometer on the board, which can be used for movement measurements. 
Depending on the headset used with it, it can be used with both dry and wet electrodes. Unfortunately, the device has no certifications and is, therefore, unsuitable for out-of-lab applications  \cite{openbci_specifications_website, consumer_grade_EEG}.

\subsubsection{Neurosky MindWave Mobile 2}

Neurosky is one of the oldest consumer-grade \gls{eeg} measurement device manufacturers. The MindWave Mobile 2 features one dry electrode and no additional sensors. Its automatic blink detection and \gls{eeg} frequency range output offer simple analysis, validated in many studies. There is an \gls{sdk} for the \gls{eeg} data to be output in a raw format or processed data as attention and meditation level. Underwent validation studies, particularly study by Maskeliunas \emph{et al.} \cite{maskeliunas2016consumer} the study evaluated attention and meditation calculated by MindWave devices, achieving only 22\% accuracy in distinguishing between focusing and relaxing subjects. While the algorithm showed limitations in separating these states, it may still have utility in assessing task difficulty or cognitive demand in specific contexts \cite{mindwave_specifications_website, consumer_grade_EEG}.

\subsubsection{InteraXon Muse}

Both devices made by Muse feature four \gls{eeg} channels and can be easily worn thanks to their dry electrodes. The newest model ``Muse S (Gen 2)'' was released in 2021 and has a longer battery life of 10 hours compared to the 5 hours from the older ``Muse 2''. Both models feature identical \gls{eeg} sensors and positions and also include a \gls{ppg} sensor, accelerometer, and gyroscope. Access to newer versions of the \gls{sdk} was restricted by the company since it is not an open-source \gls{sdk} for everyone. Consequently, the process of considering and approving requests for access to the \gls{sdk} may take a significant amount of time. \cite{consumer_grade_EEG, muse2_specifications_website, museS_specifications_website, muse_SDK_website}.

\subsubsection{Macrotellect BrainLink} 

Macrotellect Company provides BrainLink headbands in three models. \gls{eeg} sensors for all three models collect data with the same frequency. BrainLink Light has only one \gls{eeg} sensor, whereas BrainLink Dual features two \gls{eeg} sensors. The features that distinguish BrainLink Pro from the other two models are that BrainLink Pro can measure heart parameters with a \gls{ppg} sensor, and it also has a temperature sensor. Macrotellect also provides an \gls{sdk} for all three models, and raw data can be easily accessible besides some other parameters such as meditation and attention level, which their library computes 
 \cite{macrotellect_sdk_website,macrotellect_specifications_website}.

\begin{table*}[t]
\centering
\caption{Commercial head-bands specification comparison}


\resizebox{\textwidth}{!}{\begin{tabular}{lcccccl}
\toprule
\multicolumn{1}{c}{Device} & \gls{eeg} channel & Other sensors  &  Battery life{[}Hour{]} & Wireless connection & Raw data access \\ \midrule
SenzeBand 2 \cite{senzeband_specifications_website}                & 7 (250Hz)   & \gls{ppg},9-axis Accelerometer   & 6                            & Bluetooth 5.0
                 & Yes             \\
Emotiv EPOC X \cite{epoc_specifications_website}              & 14 (2048Hz)   & 9-axis Accelerometer     & 6                        & Bluetooth 5.0                 & Yes             \\
Muse 2 \cite{muse2_specifications_website}                     & 4 (256Hz)  & \gls{ppg},3-axis Accelerometer   & 5                            & Bluetooth 4.2                 & Yes             \\
OpenBCI Cyton EEG Headband \cite{openbci_specifications_website} & Customizable(250Hz) &  EMG, ECG, 3-axis Accelerometer & -                            & WiFi                 & Yes             \\
Neurosky MindWave Mobile 2 \cite{mindwave_specifications_website}  & 1 (512Hz)   & -    &6-8                              & BT/BLE  & Yes                         \\
Muse S \cite{museS_specifications_website}                     & 4 (256Hz)   & \gls{ppg}, 3-axis Accelerometer    & 10                            & Bluetooth 4.2                  & Yes               \\
BrainLink Pro \cite{macrotellect_specifications_website}                     & 1 (512Hz)   & \gls{ppg}, Temperature    & 3-4                            & Bluetooth 4.0                  & Yes               \\
    
\bottomrule
\end{tabular}}
\label{table:head_worn_table}
\end{table*}

\subsection{Wrist-worn devices}

As low-cost smartwatches become more accessible, their built-in sensors are being utilized more frequently. \gls{ppg} and acceleration sensors are built into most of them, which present a variety of physiological vital signs. There are many vital signs which can be derived from those sensors. Most commonly, these are pulse oxymetry readings~\cite{pulse_oximetry_plethysmographic}, respiratory rate~\cite{efficient_respiratory_rate, Pollreisz2020_LASCAS}, blood pressure~\cite{Taherinejad2020healthwear} and heart rate, but also arterial stiffness or hypo- and hypervolemia can be deduced ~\cite{a_review_on, wearable_medical_devices}. 
Emotion recognition can also be done using these sensors~\cite{Taherinejad2016emotions, Pollreisz2017}. There are even reports for emotion recognition solely on accelerometer data with fair accuracy~\cite{emotion_recognition_based,garcia2015automatic}. A combination of head-worn and wrist-worn acceleration data may yield even better results. Furthermore, wrist-bands' battery lifetime is an essential parameter in especially long-time studies where researchers compared common wrist-band \cite{a_feasibility_study}. 

Accelerometer data can be used to reduce movement artifacts, especially when the frequency of interest is close to the error frequency. Additionally, there are statistical algorithms that demonstrate a great improvement in accuracy
~\cite{detection_and_removal}. In the rest of this subsection, we provide a brief overview of prominent devices that are commercially accessible, along with their specifications, capabilities, and sensors.
\Cref{table:wrist_worn_table} shows a summary of modern wrist-worn devices.

\subsubsection{Empatica}

Aimed directly at researchers, both Empatica E4 and EmbracePlus are certified vital sign monitoring devices. Therefore, E4, as the old product, does not have a built-in display, which brings its battery life up to over 32 hours of continuous measurements~\cite{e4_specifications_website}. Empatica has integrated a display into EmbracePlus by improving battery consumption, making this device more user-friendly. Both devices have a \gls{ppg} sensor, an \gls{edas} sensor, a temperature sensor, and an accelerometer. Raw data can be downloaded from the device whose validity has been researched and confirmed \cite{Validity_of_the}. Empatica offers a commercial version of their watch focusing on remote, medical-grade monitoring of epilepsy patients. Using an Android SDK provided by Empatica, it is possible to receive raw data in real-time from the E4 wrist-band. In order to access data collected by Embrace, Empatica provides them on Amazon S3 servers\cite{embrace_spec}.    



\subsubsection{Apple Watch Series 4}

The Apple Watch 4 is the only device to measure electrical heart activities. The sensors include \gls{ppg},  acceleration, a gyroscope, and a barometer. The device is more accurate than other comparable products in heart rate measurements. Direct data access is impossible; the manufacturer has no official \gls{sdk}. This makes it difficult to implement custom analysis tools or connect the watch to other devices ~\cite{improving_heart_rate, watch4_specifications_website}.

\subsubsection{Polar OH1}

The Polar OH1 can be worn on the lower or upper arm, where its integrated \gls{ppg} and acceleration sensors can save up to 12 hours of continuous measurements. Raw and processed data can be extracted from the device using an open source \gls{sdk} provided by Polar \cite{polar2024sdk}. This device does not feature a display as its intended use is with a secondary device for interfacing ~\cite{HRV_and_stress, oh1_specifications_website}. 


\subsubsection{Mobovi TicWatch Pro}

While also featuring a \gls{ppg}, acceleration, gyroscope sensor, and a barometer, the TicWatch Pro manufacturer Mobovi offers support for custom data extraction tools. The device has been used in multiple publications and can have 4-8 hours of battery life while doing measurements ~\cite{improving_heart_rate, ticwatch_specifications_website, a_feasibility_study}.
\subsubsection{Garmin tactix® Charlie} Garmin produces many smartwatches with different categories as commercial products. One model of these products is tactix Charlie, which has \gls{ppg}, \gls{gps}, acceleration, and temperature sensor. Providing  Garmin Health Companion SDK by Garmin to access real-time data caused the use of Garmin in research. However, this SDK only provides processed heart rate from \gls{ppg} sensor that is not suitable for researchers that need raw \gls{ppg} data in their research ~\cite{de2023does, garmintactix_sensor_Data}.

\subsubsection{Fitbit} 
 Fitbit manufactures many smartwatches and trackers since 2009, introducing popular models such as Inspire, Versa, and the Charge series. Their battery lives range from 6 to 10 days. The Fitbit Charge 6 series features sensors for various health metrics, including \gls{ppg}, \gls{ecg}, \gls{edas}, SpO2 monitoring, as well as acceleration, ambient light, and \gls{gps}. Whereas processed data can be extracted using the \gls{api}, direct access to raw data is not possible
\cite{fitbit_spec}.

\begin{table*}[t]
\centering
\caption{Wrist-worn device comparison
}
\begin{threeparttable}
\resizebox{\textwidth}{!}{\begin{tabular}{lcclcc}
\toprule
Device & \gls{ppg} sampling rate {[}Hz{]} 
& \multicolumn{1}{c}{Accelerometer} & Other sensors                           & Wireless connection & Raw data access \\ 
\midrule
Empatica E4 \cite{e4_specifications_website}                & 64                                 & X                        & Temperature, Electrodermal Activity     & BT                  & Yes*             \\
EmbracePlus \cite{embrace_spec}                              & 64                                 & X                        & Temperature, Electrodermal Activity     & BT                  & Yes** \\
Apple Watch Series 4 \cite{watch4_specifications_website}       & -                                  & X                        & Barometer, GPS, electrical heart sensor & BT, LTE, Wi-Fi      & No              \\
Polar OH1 \cite{oh1_specifications_website}                 & -                                  & X                         & -                                       & BT                  & Yes             \\
Mobvoi TicWatch Pro 3 Ultra GPS \cite{ticwatchpro3ultra_spec}        & 100                                 & X                        & SpO2, Barometer, GPS                                     & BT, Wi-Fi            & Yes             \\
Garmin tactix® Charlie \cite{garmintactix_spec}              & -                                   & X                        & GPS, Temperature                                         & BT, Wi-Fi            & No             \\
Fitbit \cite{fitbit_spec}                                    & -                                   & X                        & \gls{ecg}, SpO2, \gls{gps}, Ambient Light                & BT                  & No             \\
\bottomrule
\end{tabular}}
\begin{tablenotes}
\tiny
\item * As of late 2023, Empatica plans to shut down E4 wristband on August 2024.
\item ** Raw data are not available in real-time, but can be exported from Empatica cloud server.
\end{tablenotes}
\end{threeparttable}
\label{table:wrist_worn_table}
\end{table*}

\subsection{Finger-worn devices}
With the advancement of science and electronic components, the size of sensors has decreased, and they can be embedded in the smallest wearable devices.
In this way, smart rings equipped with various sensors have been produced and are available to everyone to receive vital signs and monitor their health. Smart rings like smartwatches can be employed to perform mid or long-term health monitoring because they are easy to use.

One of the advantages of this wearable device is that it can be easily used everywhere with its beautiful design, long battery life, and waterproofness. 
{
However, users can download the processed data from the cloud. For none of the following finger-worn devices is access to raw data possible
\cite{oura_ring_website,motiv_specifications_website,cart_specifications_website}.}
\Cref{table:finger_worn_table} shows a summary of modern finger-worn devices. 

\subsubsection{Oura ring} One of the smart rings introduced in 2015, now with its third generation available in the market, is called Oura.
 Having a 3D accelerometer, SPO2, temperature, and \gls{ppg} sensors has given this smart ring many capabilities. Owing to these capabilities and having a strong battery that, on average, a fully charged ring will last 4-7 days, the researchers were attracted to use this smart ring in their research. One of the applications for which researchers have used the ring is sleep monitoring. It can be used to find total sleep time and sleep efficiency. In addition, with temperature sensors, researchers used the Oura ring for fever monitoring. There is a dedicated \gls{sdk} for processed data extraction and analysis from the manufacturer ~\cite{oura_ring_website, sleep_tracking_oura, fever_tracking_oura}.

\subsubsection{Motiv} 
Motiv ring is a ring designed for monitoring daily activities and also monitoring sleep. With this feature, users can look at their sleep duration and metrics like sleep restlessness and resting heart rate through the Motiv app. These features exist because a 3-axis accelerometer and a \gls{ppg} sensor are embedded in the Motiv ring. Researchers use this ring to process heart rate during trail running by having these abilities \cite{heart_rate_motiv}.

\subsubsection{CardioTracker (CART)} 
CART utilizes \gls{ppg} sensor data for the purpose of heart rate measurement and to identify Atrial Fibrillation (AF) or its burden, and Electrocardiogram (ECG) signals to offer supplementary information to doctors. The lack of an accelerometer sensor in this ring shows that the manufacturer just focused on detecting heart diseases and no more. Researchers evaluated the diagnostic performance of the CART ring, and it detected AF using deep learning analysis of \gls{ppg} signals \cite{cardiography_cart}.

\begin{table*}[t]
\centering
\caption{Finger-worn device comparison
}
\resizebox{\textwidth}{!}{
\begin{tabular}{lcccccc}
\toprule
\multicolumn{1}{c}{Device} &\gls{ppg} & Accelerometer & Temperature & Wireless connection & Other Sensors & Raw data access \\ 
\midrule
Oura Ring~ \cite{oura_ring_website}                  & X   & X & X                           & BLE                 & SpO2        & No     \\
Motiv~\cite{motiv_specifications_website}              &X     & X &                         & BT                 &     &  No       \\
CART~\cite{cart_specifications_website}                   & X   &  &                             & BLE                 & ECG     & No        \\
\bottomrule
\end{tabular}
}
\label{table:finger_worn_table}
\end{table*}

\subsection{Chest-worn devices}
Another category of wearable devices capable of assessing heart variability and consequently determining stress levels, along with other potential applications, includes chest-worn gadgets. These devices mostly feature \gls{ecg} sensors designed for heart rate monitoring. Examples of chest bands incorporating \gls{ecg} sensor are the Polar OH10 \cite{polar_h10_specifications_website} and Garmin HRM Dual 
\cite{garmin_hrm_dual_specifications_website}.

Although some chest-worn devices are used to measure heart rate, extracting data from those has proved very difficult. Additionally, wearing a chest belt can be annoying to the employee and has been reported to be inconvenient to wear and reduce mobility. On the contrary, chest-worn devices do not yield notably improved results compared to wrist-worn devices, such as in stress detection, which is one of the applications of these health monitoring devices\cite{pinge2022}. Therefore, this type of device is of minor practical relevance ~\cite{improving_heart_rate, ppg_sensors_a}. A photo depicting a subject engaged in stability activities for a stress detection study is depicted in Figure \ref{wearable_devices}, showcasing the subject wearing various wearable healthcare devices.\footnote{ This photo was taken with the permission of the subject at the Sport and Sports Sciences Institute (ISSW) of Heidelberg University.}

{\begin{figure}[htbp]
\centerline{\includegraphics[width=0.5\columnwidth]{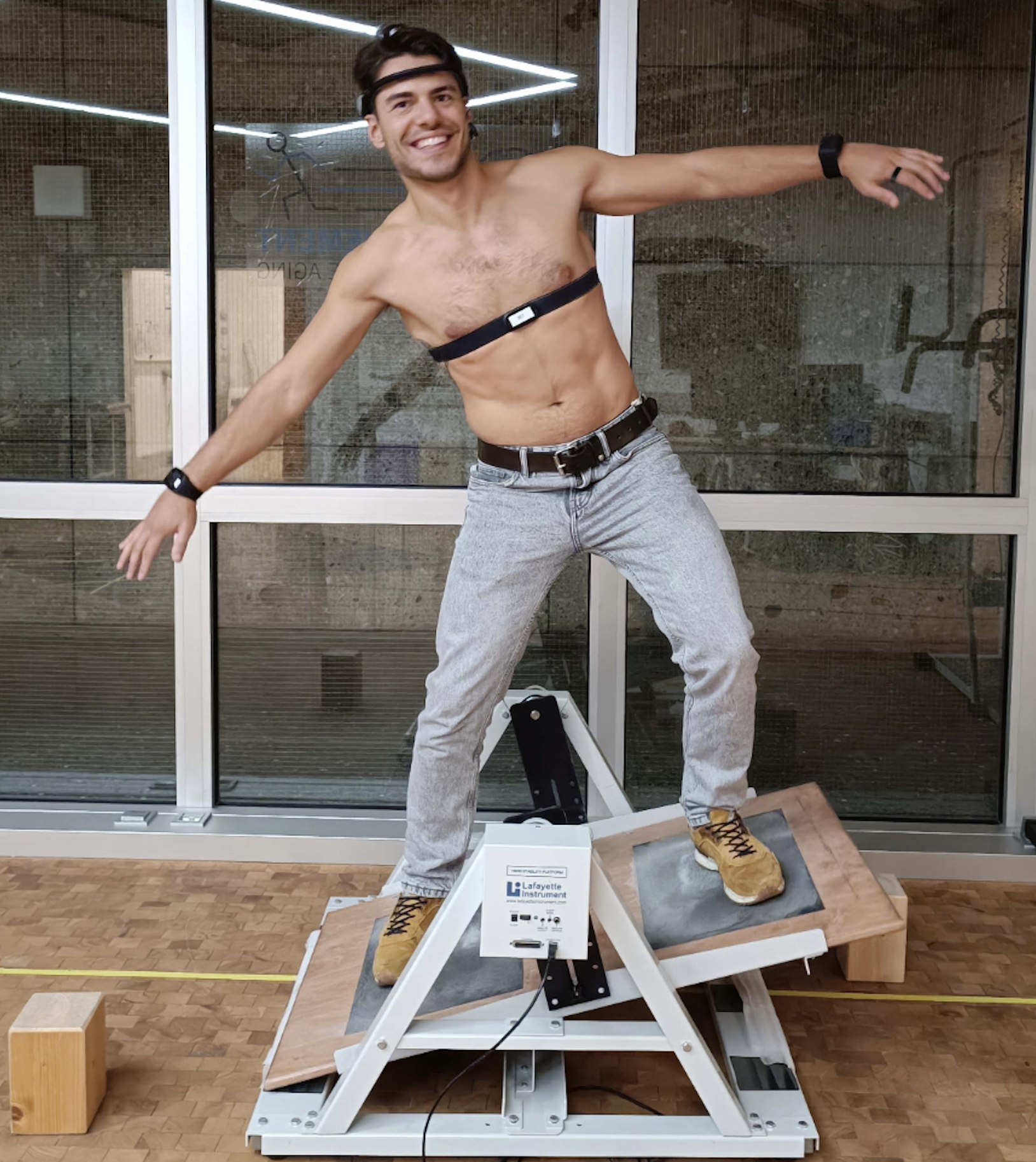}}
\caption{A Subject wearing different wearable healthcare devices including Empatica E4 and Polar M430 smartwatches, BrainLink headband, Polar chestband, and Oura Ring while performing a stress detection activity \cite{e4_specifications_website,Polar_Watch_M430,polar_h10_specifications_website,macrotellect_specifications_website,oura_ring_website}.} 
\label{wearable_devices}
\end{figure}}


\section{Applications}  \label{applications}

Raw signal data lacks meaning on its own. In order to grasp the significance of sensor data, processing is essential. Initial processing transforms raw data into meaningful, vital signs like heart rate, blood pressure, or generally \gls{ews}, which is a medical appliance used to evaluate a patient's well-being.

Employing a suitable combination of vital signs enables us to detect significant occurrences in the human body not only in the medical field, such as detecting heart diseases, but also in activity detection, education, gaming, cognition studies, neuroscience, and others ~\cite{accurate_fall_detection, dell2022machine, orlandic2021wearable,sopic2022personalized}.

Numerous retrievable vital signs can be obtained from sensors attached to the body. In order to be able to keep the applications concise, this section delves into the diverse applications of biosignals, which can be derived from the sensors mentioned in \Cref{sec_wearable_devices}.
Table \ref{table:vital_signs_table} shows a summary of vital signs derived by different sensors.

\begin{table*}[t]
\centering
\caption{Extractable vital signs per sensor, based on  ~\cite{wearable_health_devices, HRV_and_stress, consumer_grade_EEG, an_unobtrusive_and, gamma_alpha_delta}}
\resizebox{\textwidth}{!}{\begin{tabular}{rlll}
\toprule

\multicolumn{1}{l}{Sensor}                                                                              & Primary vital sign                                                                                             & Derived vital sign         & Application examples                                                       \\ \midrule
\multicolumn{1}{r|}{\multirow{7}{*}{\begin{tabular}{l}\gls{ppg}\end{tabular}}} & \multicolumn{1}{l|}{\multirow{2}{*}{\gls{hr}}}                                                                       & \multicolumn{1}{l|}{\gls{hrv}}   & Physical training intensity and duration, arrhythmia                       \\ \cline{3-4} 
\multicolumn{1}{r|}{}                                                                                   & \multicolumn{1}{l|}{}                                                                                          & \multicolumn{1}{l|}{\gls{rhr}}   & Atherosclerosis, myocardial schemia, increased apotosis                    \\ \cline{2-4} 
\multicolumn{1}{r|}{}                                                                                   & \multicolumn{1}{l|}{\multirow{3}{*}{\gls{bp}}}                                                                       & \multicolumn{1}{l|}{\gls{sbp}}   & Hypertension, cardiovascular diseases                                      \\ \cline{3-4} 
\multicolumn{1}{r|}{}                                                                                   & \multicolumn{1}{l|}{}                                                                                          & \multicolumn{1}{l|}{\gls{dbp}}   & Hypertension, cardiovascular diseases                                      \\ \cline{3-4} 
\multicolumn{1}{r|}{}                                                                                   & \multicolumn{1}{l|}{}                                                                                          & \multicolumn{1}{l|}{\gls{sbpp}}  & Vascular aging                                                             \\ \cline{2-4} 
\multicolumn{1}{r|}{}                                                                                   & \multicolumn{2}{l|}{\gls{rr}}                                                                                                                     & Sleep apnea syndrom, cardiac arrest, chronic obstructive pulmonary disease \\ \cline{2-4} 
\multicolumn{1}{r|}{}                                                                                   & \multicolumn{2}{l|}{SpO2}                                                                                                                   & Hypoxia, athletic performance, lung gas exchange diagnostics                \\ \hline
\multicolumn{1}{r|}{\multirow{5}{*}{\begin{tabular}{l}\gls{eeg}\end{tabular}}}                  & \multicolumn{1}{l|}{\multirow{5}{*}{\begin{tabular}[c]{@{}l@{}}Electrical \\ frequency \\ bands\end{tabular}}} & \multicolumn{1}{l|}{Delta} & Signal detection and decision making, sleep experiments                    \\ \cline{3-4} 
\multicolumn{1}{r|}{}                                                                                   & \multicolumn{1}{l|}{}                                                                                          & \multicolumn{1}{l|}{Theta} & Selective attention interpretation, cognition experiments                  \\ \cline{3-4} 
\multicolumn{1}{r|}{}                                                                                   & \multicolumn{1}{l|}{}                                                                                          & \multicolumn{1}{l|}{Alpha} & Working memory, long term memory                                           \\ \cline{3-4} 
\multicolumn{1}{r|}{}                                                                                   & \multicolumn{1}{l|}{}                                                                                          & \multicolumn{1}{l|}{Beta}  & Cognition experiments                                                      \\ \cline{3-4} 
\multicolumn{1}{r|}{}                                                                                   & \multicolumn{1}{l|}{}                                                                                          & \multicolumn{1}{l|}{Gamma} & Auditory, visual response measurements                                     \\ \hline
\multicolumn{1}{r|}{Accelerometer}                                                                      & \multicolumn{2}{l|}{Acceleration}                                                                                                           & Distance measurements, fall detection, signal noise reduction                                     \\ \bottomrule
\end{tabular}}
\label{table:vital_signs_table}
\end{table*}

\subsection{Early Warning Score}
\gls{ews} are a set of vital signs that are used to assess a patient's condition. By monitoring a patient's EWS, healthcare professionals can identify early signs of deterioration and take steps to prevent serious complications~\cite{gotzinger2017dataReliability}. 

In this part, we discuss vital signs that play a crucial role in calculating the \gls{ews}. These indicators can be derived from the sensors covered earlier in our discussion.
\subsubsection{\gls{hr}}

Extracting the heart rate from \gls{ppg} or \gls{ecg} can benefit users in multiple ways. \gls{rhr} can be used in medical applications to anticipate cardiovascular events. An increased \gls{rhr} can lead to atherosclerosis, myocardial ischemia, increased apoptosis, and many other problems \cite{resting_heart_rate}. \gls{hr} is also related to physical fitness, and \gls{hrvme} can be used to provide athletes with feedback concerning their schedule, duration, and intensity of training \cite{heart_rate_variability}. Extracting users' mental states and emotions can be done via \gls{hrv}, especially for stress. This has proven well-applicable, but also happiness or anger are feasible ~\cite{HRV_and_stress, Taherinejad2016emotions, Pollreisz2017}.

\subsubsection{\gls{rr}}

Knowing the \gls{rr} is often an important sign for knowing someone's health. It can indicate hypoxia and predicts cardiac arrests \cite{wearable_health_devices}. Moreover, it is an indicator to distinguish respiratory sinus arrhythmia \cite{heart_rate_variability}. \gls{rr} depends on many other pathological conditions, such as physical effort, heat/cold, cognitive load, and emotional stress \cite{the_importance_of}.

\subsubsection{\gls{spo2}}
Dual-wavelength \gls{ppg} is the most popular method for non-invasive \gls{spo2} measurements. It is a vital parameter that is almost always measured in clinical settings. Oxygenation levels of different measurement positions such as limbs and brain are used additionally in space, military applications, and stress detection\cite{wearable_health_devices, bhavani2022}. Extracting relative changes in the total hemoglobin and its oxygenated and reduced proportions may help recognize vascular complications like arterial occlusions \cite{photoplethysmography_for_blood}. In exercises, aerobic efficiency can be deduced by monitoring \gls{spo2} levels \cite{wearable_health_devices}.

\subsubsection{\gls{bp}}

Blood pressure dramatically influences cardiovascular diseases, and hypertension is a risk factor. Single \gls{bp} measurements done in clinics can be affected by different factors and do not include \gls{bp} variability. A reliable way to estimate both \gls{sbp} as well as \gls{dbp} from signals recorded using a \gls{ppg} sensor could help in the early detection of issues\cite{an_unobtrusive_and}.  \gls{ssbp} indicates vascular aging due to the higher arterial stiffness \cite{a_review_on}. 

\subsection{Neurological analysis}

\gls{eeg} measurements have a significant impact on medical brain analysis, and there is a growing availability of devices designed for individual mental assessment \cite{gamma_alpha_delta}. The measured current waveforms are correlated to neural activity, which is split into frequency ranges shown in Table \ref{table:eeg_spectrums} and explained in Section \ref{wearable_devices}. Research is done on various topics, including neurological disorders, linguistics, and developmental research \cite{oscillatory_eeg_dynamics}. Consumer-grade devices are often used to detect mental load and derive parameters like attention, vigilance, fatigue, and stress. Wireless devices are often used in brain-computer interfaces, education, or gaming \cite{consumer_grade_EEG}.

\subsection{Activity recognition}

Many modern wearable health devices feature an acceleration measurement as shown in Tables \ref{table:head_worn_table} and \ref{table:wrist_worn_table}. This feature can be used for many applications like fall detection, distance measurements, and gesture analysis ~\cite{a_feasibility_study, gesture_detection_system, demo_abstract_emeasure, jahanjoo2020detection}. In the medical field, these sensors have been used to get respiratory waveforms or activity and body posture \cite{wearable_health_devices}, detection of depression ~\cite{garcia2018depresjon, aminifar2021monitoring, aminifar2022extremely, jakobsen2020psykose} and schizophrenia [s1, s2] from mental health domain. In combination with other sensors, acceleration data is also often used for noise detection and removal 
\cite{detection_and_removal}.

\subsection{Mental Health}
The relation between motor activity data and mental health conditions such as schizophrenia and major depression is studied \cite{berle2010actigraphic, faedda2016actigraph}. In \cite{garcia2018depresjon}, the authors collect motor activity data from 23 unipolar and bipolar depressed patients and 32 non-depressed contributors using actigraph wristbands (Actiwatch, Cambridge Neurotechnology Ltd). In \cite{garcia2018depresjon, aminifar2021monitoring}, it has been shown that depression can be detected based on such data using \gls{ml}. In \cite{jakobsen2020psykose}, motor activity data from 22 patients with schizophrenia and 32 healthy control persons was collected using actigraph wristbands. In \cite{jakobsen2020psykose, aminifar2022extremely}, the authors show that \gls{ml} algorithms can detect such mental conditions based on motor activity data.

\newacronym{osa}{OSA}{Obstructive Sleep Apnea}

\subsection{Sleep Apnea} 
Single-channel \gls{ecg} signals collected by medical wearable sensors can be utilized for detecting sleep disorders. In~\cite{surrel2018online}, the authors propose a wearable, accurate, and energy-efficient system for monitoring \gls{osa}, which is an important underdiagnosed sleep disorder. They develop an efficient time-domain analysis to compute the score for \gls{osa} that considers wearable systems' energy constraints. According to their evaluation results based on the PhysioNet Apnea-ECG database \cite{Goldberger2000-nt}, their proposed system achieves up to 88.2\% classification accuracy and a battery lifetime of 46 days for continuous screening of \gls{osa}.

\subsection{Seizure Detection}
\gls{eeg} and \gls{ecg} signals collected by wearable devices can be used for the detection of epileptic seizures \cite{sopic2018glass, forooghifar2019self, forooghifar2021self}. An early warning from medical wearable devices can notify caregivers to rescue the patient. In \cite{sopic2018glass}, the authors propose e-Glass, which is a system incorporating four \gls{eeg} electrodes to detect epileptic seizures. Their system achieves a sensitivity of 93.80\% and a specificity of 93.37\%, and a battery lifetime of 2.71 days for continuous operation. In \cite{forooghifar2019self}, the authors propose a self-aware wearable system for real-time seizure detection based on \gls{ecg} signals. They employ \gls{ml} to detect seizures by analyzing cardiac and respiratory responses to seizures obtained from \gls{ecg} signals and introduce the notion of self-awareness in their system to ensure energy efficiency. Their system achieves from 85.54\% to 79.33\% geometric mean of specificity and sensitivity and a battery lifetime between 67.55 and 136.91 days.

\subsection{Stress detection} \label{stress_detection}
The term stress is nowadays widely understood in its meaning, defined by Hans Selye in 1936 as “the non-specific response of the body to any demand” \cite{hans_selye_and}. Short-term stress can be helpful for athletes and physically demanding tasks, but also enhance cognitive capabilities ~\cite{stress_monitoring_and, consumer_grade_EEG}. Long-term stress is detrimental to health and can lead to disorders and health problems \cite{stress_monitoring_and}.

Many different sensors and algorithms can be used for stress detection with heavily varying accuracy, wearability, and use cases. The most promising approaches for workplace environments with the devices mentioned in Section \ref{wearable_devices} and extractable vital signs mentioned in Section \ref{applications} are discussed in this section. Enhancing classification accuracy can be achieved by combining the measured parameters from diverse devices. This can also be advantageous for increasing single parameter accuracies by using a secondary measurement during their calculation, e.g., acceleration for \gls{ppg} signals \cite{combining_and_comparing}. 

Due to the complexity and high volume of  available data, various \gls{ml} techniques are employed for this purpose, and comparisons are often made based on their outcomes.~\cite{classification_of_perceived, eliminating_individual_bias, a_review_on_mental}.

\subsubsection{Personalized Stress Monitoring Using Wearable Sensors in Everyday Settings}
The work of Tazarv \emph{et al.} concludes that by only using data extracted from a \gls{ppg} sensor, binary stress detection can be done with an accuracy of up to 76\%. With a sampling rate of $20Hz$, the deduced \gls{hr} and \gls{hrv} are used as input for different ML methods. Fourteen subjects were recorded during their daily activities in an uncontrolled environment, and the measured data were labeled using self-reports \cite{personalized_stress_monitoring}. 

\subsubsection{Personal Stress-level Clustering and Decision-level Smoothing to Enhance the Performance of Ambulatory Stress Detection with Smartwatches}
Three different stress levels were done by Can \emph{et al.} with the use of two different wrist-worn devices, \gls{edas}, \gls{hr}, \gls{hrv}, skin temperature, and acceleration were extracted from 32 participants. The data was validated with questionnaires and, again, different \gls{ml} algorithms compared to each other. By shifting the results into binary classes, an accuracy of up to 92\% was achieved
\cite{personal_stress_level_clustering}.



\subsubsection{Stress: Towards a Gold Standard for Continuous Stress Assessment in the Mobile Environment}
Up to 72\% accuracy was achieved with data collected in a natural environment done by Hovsepian \emph{et al.} 21 participants were using devices for one week, and ECG, acceleration, and plethysmography data was collected. The collected data was joined with measurements from a lab-controlled study. Self-reported stress levels were collected with which an ML algorithm was trained
\cite{cstress_towards_a}.

\subsubsection{Classification of Perceived Human Stress using Physiological Signals}
By using both a head-worn and wrist-worn device, Arsalan \emph{et al.} were able to implement \gls{eeg} as well as \gls{ppg} and \gls{edas} measurements. Four statistical features were extracted from the recorded data, and stress was classified in binary with three ML classifiers. The highest classification accuracy achieved with data from 28 subjects was 75\%. The subjective data was gathered via a perceived stress scale from questionnaires before the in-lab measurements were taken \cite{classification_of_perceived}.

\subsubsection{Eliminating Individual Bias to Improve Stress Detection from Multimodal Physiological Data}
A comparison of sensors and sensor groups done by Das \emph{et al.} concludes that only using \gls{eeg} (64\%) performs worse than \gls{eeg}, \gls{edas} and \gls{ppg} combined (69\%). Also, different techniques were used and compared before the classification via an ML algorithm. The dataset was taken from 10 subjects during trials in a laboratory, and self-reported data were used in the classification. The authors split up different accompanying effects of stress into separated scales for finer incrementation and analysis \cite{eliminating_individual_bias}.

\subsubsection{Multi-Modal Acute Stress Recognition Using Off-The-Shelf Wearable Devices}
In \cite{montesinos2019multi}, the authors investigate the possibility of detecting stress based on off-the-shelf wearable devices, i.e., Shimmer3 ECG Unit and Empatica E4 wristband. First, they collect physiological signals, including \gls{ecg}, \gls{edas}, \gls{st}, and \gls{rsp}, from 30 subjects (between 25-35 years old) in three stages/states. In the first and second stages, scenery clips and horror clips, respectively, are shown to the subject, while in the third stage, an arithmetic task is given to the subject. After data collection, the signals are processed, and features are extracted. Then, the extracted features are used to train several \gls{ml} models, based on \gls{detr} \cite{quinlan1986induction}, \gls{knn} \cite{guo2003knn}, and \gls{rf} \cite{ho1995random} algorithms, for classifying the signals and detecting the states. They achieved up to 84.13\% accuracy in detecting acute stress episodes.

 \subsubsection{CAFS: Cost-Aware Features Selection Method for Multimodal Stress Monitoring on Wearable Devices}
 In \cite{momeni2021cafs}, the authors focus on the trade-off between prediction performance and energy consumption in the context of stress detection using wearable devices. They propose \gls{cafs} to find a balance between prediction-power and energy-cost for multimodal stress monitoring. \gls{cafs} selects the most important features considering constraints for energy consumption. Subsequently, they present a self-aware stress monitoring framework that intelligently transitions between the energy models to reduce energy consumption. Their framework saves energy by a factor of 10 and achieves an accuracy of 88.72\% on unseen data.

 \subsubsection{ReLearn: A Robust Machine Learning Framework in Presence of Missing Data for Multimodal Stress Detection from Physiological Signals}
 One of the significant challenges in using physiological signals recorded by conventional monitoring devices is dealing with missing data. This can happen when the contact of sensors is insufficient or when other equipment interferes in the process of collecting data. In particular, if the subject is more active or is stressed, he/she makes more conscious or subconscious movements, which can lead to missing more values. In \cite{iranfar2021relearn}, the authors address this problem and propose ReLearn, which is a robust machine learning framework for detecting stress based on biomarkers extracted from multimodal physiological signals. ReLearn handles missing data and outliers for both the training and inference phases. Their framework achieves 78\% accuracy in the presence of missing values.

\subsubsection{Comparison of Stress Detection through \gls{ecg} and \gls{ppg} signals using a Random Forest-based Algorithm}
In order to enhance the accuracy of stress detection, it is prudent to leverage all available resources. There is potential value in utilizing a model trained on \gls{ppg} data to detect stress from  \gls{ecg} data or vice-versa. In line with this, Benchekroun \emph{et al.} \cite{benchenkrun2022} not only conducted a comparison of stress detection outcomes using \gls{ecg} and \gls{ppg} sensor data, but they also experimented with an \gls{rf} model trained on \gls{ppg} data and tested on \gls{ecg} sensor data. Their model predicts stress with 82\% and 83\% accuracy when they train and test it with \gls{ppg} and \gls{ecg} sensor data, respectively. 
They also attained a 73\% accuracy rate when testing \gls{ecg} sensor data using a model trained with \gls{ppg} sensor data. This method acknowledges the possibility of effectively applying models and data sources across different applications for accurate stress detection.

\subsubsection{Stress Detection With Single PPG Sensor by Orchestrating Multiple Denoising and Peak-Detecting Methods}
Finding peaks is one of the most important phases during processing \gls{ppg} sensor data because  \gls{ppg} features for general stress detection are based on \gls{hrv}, which is quantified as the changes in the interval between successive peaks \cite{shaffer2017overview}. Indeed, precisely detecting peaks plays a crucial role in accurately detecting stress.

Heo \textit{et al.} \cite{seongsil2021} present an ensemble-based peak-detecting method that consists of combining five peak-detection methods. They apply the local maxima method, block generation with the mean of the signal threshold method, first derivative with an adaptive threshold method,  slope sum function with an adaptive threshold method, and moving averages with the dynamic threshold method. Ultimately, they determine the final peak point through a majority voting process among the five points. Ultimately, they achieved a classification accuracy of 95.07\%. 

\subsubsection{High-Accuracy Stress Detection Based on \gls{ppg} Sensor Embedded in Smartwatches}
In \cite{jahanjoo2024high}, the authors propose a stress detection method with a single \gls{ppg} sensor integrated into smartwatches. They employ different time-domain, frequency-domain, and nonlinear-domain \gls{hrv} features to train several \gls{ml} models. To find the best signal window sizes for detecting stress, they calculate the results for different window sizes, determining that a size of 360s yields the best outcomes. Additionally, they explore different denoising techniques including bandpass, Kalman, and moving average filtering to enhance the signal's quality. They achieve 95.55\% accuracy and 91.42\% F1-score using the \gls{svm} \cite{cortez1995support} algorithm.

A summary of the various approaches employed for stress detection is presented in Table \ref{tab:stress_decetion_works}. It includes information on the signals utilized for stress detection, the different machine learning algorithms applied, the highest achieved accuracies, and their primary contributions or novelties.

\begin{table}[!ht]

\caption{Summary of key findings in stress detection studies}

\centering
\small
\begin{tabular}{p{0.75cm} p{2.2cm} p{3cm} p{0.5cm} p{1.6cm} p{5.4cm}}
\toprule
\textbf{Paper} & \textbf{Biosignal} & \quad\quad\textbf{Classifier} & \textbf{Acc.} & \textbf{Subjects \#} & \quad\textbf{Major contribution/Novelty} \\
\midrule
 \cite{jahanjoo2024high} & \gls{ppg} & Ada-Boosting \cite{freund1997decision}, \gls{knn}, \gls{lda} \cite{balakrishnama1998linear}, \gls{rf}, \gls{svm}, 
 \gls{dtc} \cite{quinlan1986induction} &  95\% & 15 \cite{schmidt2018introducing} & Extracting time-domain, frequency-domain, and nonlinear-domain \gls{hrv} features withing \gls{ppg} signals for detecting stress\\
\hline
 \cite{seongsil2021} & \gls{ppg} & \gls{dtc}, \gls{rf}, Ada-boosting, \gls{knn}, \gls{lda} , \gls{svm}, and Gradient-boosting \cite{friedman2001greedy} &  95\% & 15 \cite{schmidt2018introducing} & Using two-step denoising and an ensemble-based multiple peak-detection method for improving stress detection performance\\
 \hline

 \cite{personal_stress_level_clustering} & \gls{ppg}, \gls{eda}, \gls{st}, and Acceleration & \gls{mlp} \cite{popescu2009multilayer}, \gls{svm}, \gls{knn}, \gls{rf}, and \gls{lda} & 92\% & \quad32 & Developing a hybrid person-independent stress detection model without requiring an extensive personal data collection period\\
\hline

\cite{momeni2021cafs} & \gls{ecg}, \gls{rsp}, \gls{ppg}, \gls{eda}, and \gls{st} & \gls{xgb} ~\cite{chen2016xgboost} & 91\% & \quad 60 &  Using a cost-aware feature selection approach to trade off between prediction-power and energy-cost \\
\hline

 \cite{montesinos2019multi} & \gls{ecg}, \gls{eda}, \gls{st}, and Respiration & \gls{dtc}, \gls{knn}, and \gls{rf} &  84\% &\quad30 &  Demonstrating the feasibility of recognizing acute stress using off-the-shelf wearable devices and multi-modal machine learning technique\\
\hline

 \cite{benchenkrun2022} & \gls{ecg}, and \gls{ppg} & \gls{rf} & 83\% &\quad46 &  Demonstrating that \gls{hrvme} features extracted from \gls{ppg} signal serve as reliable alternatives to those extracted from \gls{ecg} signal \\
\hline

 \cite{iranfar2021relearn} & \gls{ecg}, \gls{rsp}, \gls{ppg}, and \gls{eda} & \gls{xgb} &  78\%  & \quad95 \cite{rodrigues2020locomotion} & Overcoming challenges of missing data and outliers using multi-modal physiological signals \\
\hline

 \cite{personalized_stress_monitoring} & \gls{ppg} & \gls{mlp}, \gls{svm}, \gls{knn}, \gls{rf}, and \gls{xgb} & 76\% &\quad14 &  Exploring the objective prediction of stress levels in everyday settings through a layered system architecture for personalized stress monitoring, using low-cost \gls{ppg} sensors available on common wearable devices\\
\hline

 \cite{classification_of_perceived} & \gls{eeg}, \gls{ppg}, and \gls{eda} & \gls{svm}, \gls{mlp}, and \gls{nb} & 75\% & \quad28 &  Introducing a novel approach to classifying perceived human stress using non-invasive physiological signals (\gls{eeg}, \gls{eda}, \gls{ppg})\\
\hline

 \cite{cstress_towards_a} & \gls{ecg}, \gls{ppg}, and Acceleration & \gls{svm} & 72\% &\quad  21 & Addressing the lack of a gold standard by undergoing comprehensive computational modeling steps, including data collection, screening, cleaning, filtering, feature computation, normalization, and model training\\
\hline

 \cite{eliminating_individual_bias} & \gls{ecg}, \gls{ppg}, and \gls{eda} & \gls{rf} & 69\%  &\quad 10 (146 instances) ~\cite{reuderink2009affective} &  Eliminating the subject-specific bias for classification accuracy improvement\\
\bottomrule
\end{tabular}
\label{tab:stress_decetion_works}
\end{table}

\subsection{Attention detection}\label{attention_detection}

In a lot of workplace environments, the level of attention of a single person is crucial, especially where single errors e.g. those made by pilots or police officers, can lead to fatal accidents \cite{photoplethysmographic_waveform_versus}. The attention level varies during the day and is influenced by many factors such as sleep quality, food intake, or exercise. Monitoring this and indicating when certain thresholds have been crossed allows employees to take breaks when needed. This in turn makes the employee more productive and ensures a good work-life balance \cite{wearable_ppg_sensor}. 

During an academic course, the process of assessing material can be expedited when the level of attention among the students is high. If breaks are timed accordingly, both the perceived as well as the actual speed of learning is higher \cite{learning_immersion_assessment}. 

Most techniques use the data from head-worn \gls{eeg} or eye-tracking devices. \gls{ppg} and other vital sign measurements are often added, but only a few methods solely rely on wrist-worn devices \cite{wearable_ppg_sensor}.

\subsubsection{Human Attention Recognition with Machine Learning From Brain-EEG Signals}
Hassan \emph{et al.} conducted a study involving 30 participants and categorized the attention level into three distinct groups using \gls{eeg} data. Each phase of attention was normally between 1-2 seconds, followed by one of the three other neutral, happy, or boring states. The data was collected in a controlled environment. After training an \gls{ml} algorithm and subsequently solving mathematical problems, accuracy of almost 89\% was achieved
\cite{human_attention_recognition}.

\subsubsection{Classification of Human Attention to Multimedia Lecture}
To precisely classify states of attention, Lee \emph{et al.} recorded \gls{eeg} data from eight participants during the process of learning. With this data, a binary classification accuracy of 72\% was reached. The used \gls{eeg} electrode positions are compatible with mobile head-worn devices
\cite{classification_of_human}.

\subsubsection{Wearable \gls{ppg} Sensor-Based Alertness Scoring System}
Due to the rich source of data available for ECG and the high correlation between RR series extracted from both \gls{ppg} and \gls{ecg} signals, Dey \emph{et al.} opted to initially train an \gls{svm} \cite{cortes1995support} model using features extracted from \gls{ecg} RR series, despite relying solely on data recorded from a \gls{ppg} sensor in the application. The data was sampled continuously in an uncontrolled environment. The achieved accuracy for awake or sleep classification was over 80\%, based on the participant profile
\cite{wearable_ppg_sensor}.


\subsubsection{Photoplethysmographic Waveform Versus Heart Rate Variability to Identify Low-Stress States: Attention Test}
ECG and \gls{ppg} data recorded from 50 participants were used to determine attention levels. The states were defined as resting and attention and tested in laboratory conditions. By using parameters extracted from \gls{ppg} only, an accuracy of 89\% for binary classification was achieved, while using ECG yielded worse results
\cite{photoplethysmographic_waveform_versus}. 

\subsubsection{Evaluation of Learning Performance by Quantifying User’s Engagement}
By measuring \gls{eeg}, eye movement, \gls{ppg},  and \gls{eda} during sessions,  Sandhu \emph{et al.} were able to reach an accuracy of 90\%. The attention level was split into three classification levels, and ten participants were tested in controlled environments. The data analysis was done on single sensor data only, and no combined data was used. The best results were reached with the \gls{eeg} data 
\cite{evaluation_of_learning}.

\subsubsection{Attention Detection using Electro-oculography Signals in E-learning Environment.
}
This research paper addresses the challenge of maintaining student focus during online lectures or videos. Abdo \emph{et al.} in \cite{Abdo2021} propose an attention detection system utilizing \gls{eog} signals generated by eye movements. They classify six types of eye movements and collect a substantial dataset from 50 subjects. Signal noise is filtered using a band-pass filter, and deep learning models, including \gls{cnn} \cite{albawi2017understanding} and Inception network, are employed for classification. Notably, the inception model achieved an impressive average accuracy of 93.63\% for attention detection.

\subsubsection{Frequency-based EEG Human Concentration Detection System Methods with SVM Classification}
In \cite{Purnamasari2019}, authors focus on recognizing human concentration, particularly crucial in activities like driving to prevent accidents. \gls{eeg} signals from a Neurosky Mindwave headband are employed for this purpose. The study compares two frequency-based feature extraction methods: \gls{psd} from \gls{fft} and energy from \gls{dwt}. \gls{dwt} outperforms \gls{fft}, improving accuracy by 18\%. Using a \gls{svm} classifier the system achieves a 91\% accuracy in detecting human attention.

\subsubsection{Classification of Relaxation and Concentration Mental States with \gls{eeg}}
Scientists have demonstrated the feasibility of detecting concentration using a single \gls{eeg} sensor device. 
In \cite{you2021classification}, similar to previous work, the author investigates detecting concentration using a cost-effective Neurosky Mindwave device with a single \gls{eeg} channel. Subjects are asked to recite numbers backward while their \gls{eeg} signals are converted into time-frequency representations. \gls{svm} and multi-layer feed-forward networks are employed as classifiers, individually trained for each subject. Results demonstrate that, with features from alpha, beta, and gamma bands with a 4 Hz bandwidth, the average accuracy of detecting concentration level exceeds 80\% across subjects. 

\subsubsection{Classification of Human Concentration in EEG Signals Using Hilbert Huang Transform}
One of the most important steps in concentration detection using \gls{eeg} sensor data is feature extraction techniques. 
In \cite{aziz2017classification}, the authors explore the use of \gls{hht} to classify concentration states based on EEG signals. \gls{hht} involves two steps: \gls{emd} and Hilbert Transform. \gls{emd} decomposes the \gls{eeg} signal into \gls{imf}, while Hilbert Transform is used to calculate the power spectrum for feature extraction in different \gls{eeg} signal bands. 
They employ the Extreme Learning Machine (ELM) as the classifier. 
Comparing \gls{hht} with single \gls{imf} to the \gls{fft} method, results indicate an improvement and they detect concentration with 72\% accuracy.

\subsubsection{Driver Vigilance Estimation with Bayesian LSTM Auto-encoder and XGBoost Using EEG/EOG Data}
Detecting attention, especially in applications like monitoring a driver's attention to prevent distractions, has major importance. Such detection can potentially save lives by triggering timely alarm systems, ensuring the safety of the driver and others on the road.
In \cite{zeghlache2022driver}, the authors address driver drowsiness detection using machine learning applied to \gls{eeg} and \gls{eog} data. They present two \gls{vae} models to estimate driver alertness by reducing \gls{eeg} features while maintaining classifier accuracy. Using a multimodal dataset, \gls{eeg} and \gls{eog} features and applied to an XGBoost classifier, with the \gls{vae} LSTM encoder offering improved feature extraction. This approach achieved 76\% for precision. 

A summary of the various approaches employed for attention detection is presented in Table \ref{tab:attention_decetion_works}. It includes information on the signals utilized for attention detection, the different machine learning algorithms applied, the highest achieved accuracies, and their primary contributions or novelties.

\begin{table}[!ht]
\caption{Summary of key findings in attention detection studies}
\centering
\small

\begin{tabular}{p{1cm} p{2cm} p{3cm} p{0.5cm} p{1.5cm} p{5cm}}

\toprule
\textbf{Paper} & \textbf{Biosignal} & \quad\quad\textbf{Classifier} & \textbf{Acc.} & \textbf{Subjects \#} & \quad\textbf{Major contribution/Novelty} \\
\midrule
\cite{Abdo2021} & \gls{eog} & \gls{cnn}, \gls{vgg}, and Inception &  94\%  &\quad50& Introducing an attention detection system using \gls{eog} signals to identify non-attention periods and recognize \gls{adhd} patterns in students during learning\\
\hline
 \cite{Purnamasari2019} & \gls{eeg} & \gls{svm} & 91\% & \quad10 &  Introducing a method for recognizing human concentration using a \gls{bci} system employing frequency-based feature extraction\\
\hline
 \cite{evaluation_of_learning} & \gls{eeg}, \gls{ppg}, \gls{eda}, and Eye Tracker  & \gls{svm} & 90\% & \quad10 & Employing multi-modal sensors to objectively quantify engagement levels in real-time learning activities\\
\hline

 \cite{human_attention_recognition} & \gls{eeg} &\gls{cnn}-LSTM &  89\% &\quad30 &  Introducing a cost-effective single-channel \gls{eeg}-based attention level recognition system, using diverse datasets and a hybrid \gls{cnn}-LSTM model\\
\hline
 \cite{photoplethysmographic_waveform_versus} & \gls{ppg}, and \gls{ecg} & Bagging using \gls{dtc} & 89\% & \quad50&Developing an automatic identifier for attentional states, by extracting the most appropriate features for detecting a subject's high performance state \\
\hline
 \cite{you2021classification} & \gls{eeg} & \gls{svm}, and \gls{mlp} & 84\% &1\textsuperscript{st} run: 7, 2\textsuperscript{nd} run: 10 & Presenting a study based on one-channel toy-grade \gls{eeg} device to classify concentrated and relaxed mental states\\
\hline

 \cite{wearable_ppg_sensor} & \gls{ppg}, and \gls{ecg} & \gls{svm} & 80\% & \quad- & Proposing a smartwatch-based system using \gls{ppg} and \gls{ml} to continuously estimate alertness levels \\
 \hline

 \cite{zeghlache2022driver} & \gls{eeg}, and \gls{eog}& \gls{xgb} & 76\%& \quad23 trials (each trial 2 hours) ~\cite{zheng2017multimodal}  & Presenting an approach for driver drowsiness detection using two \gls{vae} models, aimed at reducing the dimensionality of \gls{eeg} features while maintaining a high level of accuracy\\
\hline
 \cite{classification_of_human} & \gls{eeg} & \gls{knn}, \gls{svm}, \gls{cnn}, and Ensemble &  72\% &\quad 8 & Classification of attention and non-attention to online lecture videos by extracting \gls{psd} features from \gls{eeg} signals\\
\hline

 \cite{aziz2017classification} & \gls{eeg} & \gls{elm} \cite{huang2006extreme} & 72\%&\quad 5 & Introducing an approach for classifying concentration states using \gls{hht} \\
\bottomrule
\end{tabular}
\label{tab:attention_decetion_works}
\end{table}

\section{Discussion}
\label{sec_discussion}
\begin{figure}%
    \centering
    \subfloat[\centering Attention Detection]{{\includegraphics[width=0.45\columnwidth]{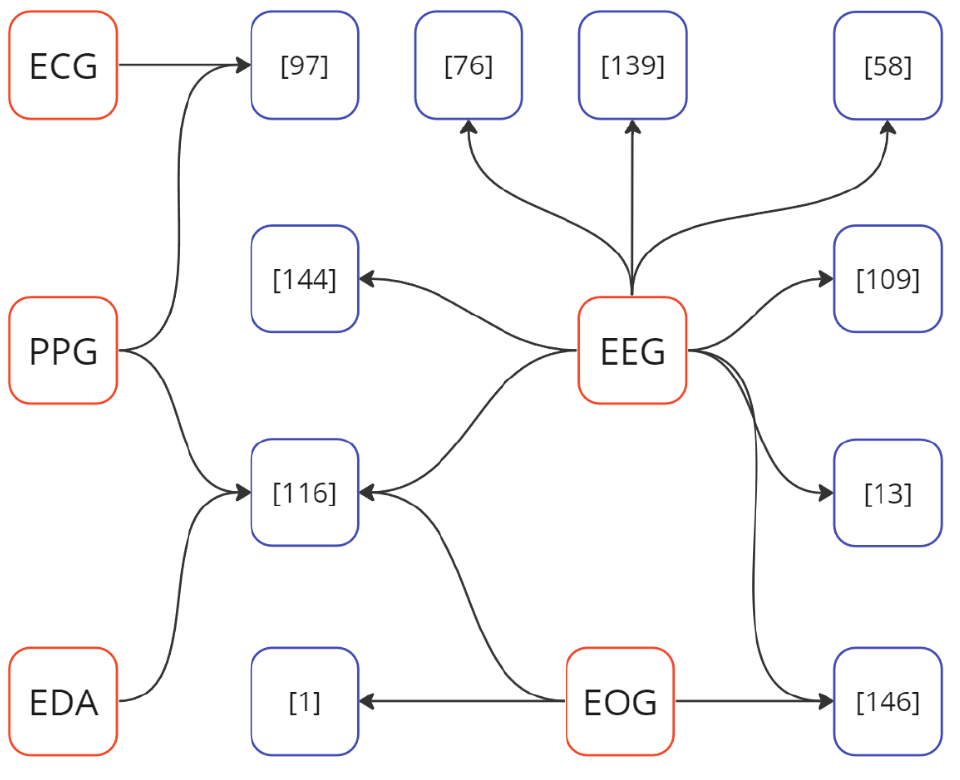} }}%
    \qquad
    \subfloat[\centering Stress Detection]{{\includegraphics[width=0.43\columnwidth]{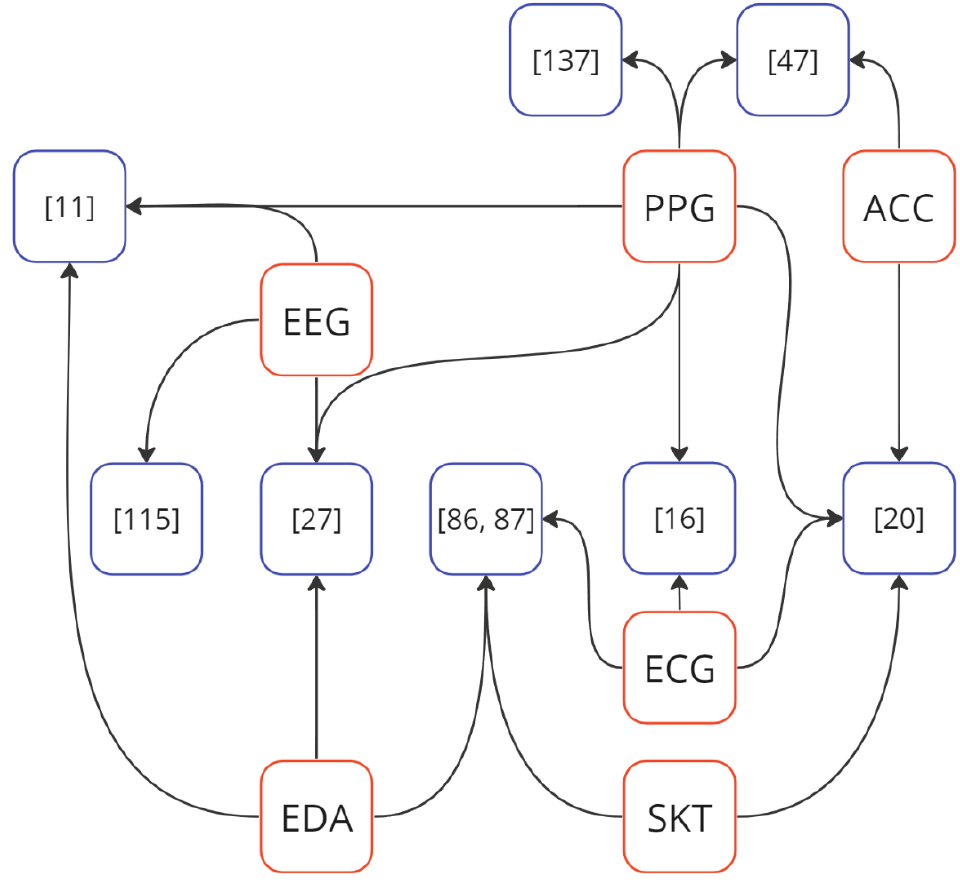} }}%
    \caption{Overview of biosensor usage in stress and attention detection research (SKT: Skin temperature, EOG: Electrooculography)}%
    \label{fig:attention_stress}%
\end{figure}

\begin{table*}[t]
\centering
\caption{Used ML algorithms for emotion detection. In this table, MLP=Multi Layer Perceptron, RF=Random Forest, kNN=k-Nearest Neighbors, SVM=Support Vector Machine, LDA=Linear discriminant analysis, NB=Naive Bayes, CNN-LSTM=Hybrid Convolutional neural network - Long short-term memory.}
\begin{tabular}{l|ccccc|ccccc}
\toprule
 Machine learning & \multicolumn{5}{c|}{Top stress detection accuracy} & \multicolumn{5}{c}{Top attention detection accuracy} \\
\multicolumn{1}{c|}{classification algorithm} & \cite{personalized_stress_monitoring} & \cite{personal_stress_level_clustering} & \cite{classification_of_perceived} & \cite{momeni2021cafs} & 
\cite{seongsil2021}&
\cite{classification_of_human} & \cite{wearable_ppg_sensor} & \cite{human_attention_recognition} & \cite{evaluation_of_learning}&
\cite{you2021classification}\\ \cline{1-11}
SVM      & 69\%  & 70\%  & 50\%  &       &94\%       & 76\%  & 80\%    &       & 90\%  &84\% \\
CNN-LSTM &       &       &       &       &       &       &        & 89\%  &      &  \\
MLP      & 73\%  & 70\%  & 75\%   &      &       &       &         &       &      &82\%  \\
kNN      & 72\%  & 86\%  &       &      &92\%       & 77\%  &         &       &     &   \\
LDA      &       & 62\%  &        &       &95\%      &       &         &       &     &   \\

Ensemble &   76\%    &  92\%     &       &  91\%     & 91\%     & 72\%  &         &       &       &    \\
\bottomrule
\end{tabular}
\label{table:ML_algos_table}
\end{table*}

In this research, our primary emphasis was dedicated to the exploration and explanation of healthcare-oriented wearable devices, containing a collection of headbands, wristbands, smart rings, and chestbands that have been equipped with specialized sensors to cover extensive healthcare monitoring applications. 
Since how (i.e., where) such devices are worn plays an important role in their utility, we presented a first categorization approach based on different device positions, 
i.e., wrist, head, and finger. We explained the details of these devices and the specifications related to sensor settings, connection protocols, and raw data access. The second categorization takes a different approach by focusing on end applications. We studied several applications where wearable devices were used and discussed what they achieved and how.

Table \ref{table:head_worn_table} shows that almost all head-worn devices feature accelerometers, and half have integrated \gls{ppg} sensors. This table presents that sampling rate as one of the specifications of \gls{eeg} sensor, can vary from 250 to 2048~Hz. 
The range of different frequency bands of  interest in the \gls{eeg} signal is from delta to gamma, i.e., between 0.5 and 80~Hz~\cite{jadeja2021}, and according to the Nyquest-Shanon sampling theorem \cite{landau1967}, any signal with limited energy and $f$~Hz bandwidth may be reconstructed 
via samples taken with a bandwidth of $2f$~Hz or higher. Consequently, the sampling rate in the range of few hundreds is sufficient for applications that use the gamma as the maximum frequency range of interest.
Higher sampling frequency rates 
require more storage space and processing power to handle the increased volume of the generated data. 
Considering that the main sensor of head-worn devices is \gls{eeg}, these devices are mostly used to monitor brain-related activities such as epileptic seizure or cognitive processes. In this article, we investigated the applications of this sensor in the detection of attention as a cognitive process, which can have various applications, such as detecting attention while working or participating in an online course.

 Chest-worn devices can provide accurate and reliable heart rate data, which can be helpful for monitoring cardiac health, fitness, and performance. However, they are often inconvenient, may reduce mobility, and are not ideally suited for a long time wearing~\cite{improving_heart_rate}. Therefore, they are usually used for relatively shorter term monitoring, especially when higher precision is required. 
 
Table \ref{table:wrist_worn_table} presents wrist-worn devices, often smartwatches, usually equipped with \gls{ppg} and accelerometer sensors to monitor vital signs and physical activities. Acceleration data are useful for detecting noise in \gls{ppg} signal too. Other sensors like \gls{eda} and temperature are mostly embedded in the devices that are produced for research purposes. 
 In this table, we present a simple device called the Polar OH1, which solely incorporates a \gls{ppg} sensor, as well as more complex devices like the EmbracePlus, which includes additional sensors such as \gls{eda} and accelerometer alongside the \gls{ppg}. 
By having these sensors, wristbands can provide data to measure various health parameters such as heart rate, blood pressure, and oxygen saturation. 
Additionally, we can track physical activity, monitor sleep quality, and detect stress levels by processing the sensors data.
However, it is important to note that the data they provide are not as accurate and reliable as clinical data due to different sources of noise during data collection \cite{detection_and_removal,aminifar2024recognoise}.
However, they are easy to use and convenient to wear, hence very suitable for longer period monitoring, where clinical-level precision is not necessary. 

In Figure \ref{sensor_to_application}, we demonstrated the sensors that 
are integrated within each distinct category of devices, such as headbands, wristbands, and smart rings based on devices available in the market. This overview serves to equip other researchers with an insightful perspective, enabling them to distinguish and select the most suitable device category based on their specific sensor requirements and operational needs and preferences. More importantly, it provides a clear overview of their overlap, where any of those wearable devices could be used (almost) interchangeably or serve as redundancy one for the other.



Given the limited range of feasible wearing positions, such as the head and wrist, numerous devices feature similar sensors~\cite{HRV_and_stress, a_review_on, consumer_grade_EEG}. \gls{ppg} being the most common one, whereas \gls{ecg} and \gls{eda} were quite common too. 
\gls{eeg} has been used in many attention and stress detection works. In Figure \ref{fig:attention_stress} (a), we showcase the studies and the sensors they utilized in various works to detect attention. Since attention is commonly derived from processing data channels of \gls{eeg} sensor, we observe that most of the research applied this sensor to detect attention. However, in \cite{photoplethysmographic_waveform_versus}, researchers demonstrate that it is possible to detect attention using \gls{ppg} sensor data instead of relying on \gls{eeg}.
This can open a new field of research and potentially many new use cases where headworn devices are not practical. 

Figure \ref{fig:attention_stress} (b) illustrates the set of sensors used in each research for detecting stress. Unlike attention detection, which is mainly computed by EEG sensors, typically headworn devices with minimal movement, most devices used for stress detection are worn on body parts that 
are subject to significant movement, such as the wrist. This movement often introduces significant noise. Certain studies depicted in this figure utilize accelerometers to monitor and remove the noise resulting from this movement in the physiological signals.



The measured physiological parameters can be combined to achieve higher overall accuracy. However selecting the appropriate ones and determining the optimal combination, is often a challenging task, as Table \ref{tab:stress_decetion_works} shows many research utilized different sensors to achieve a high accuracy but the study that achieved the highest accuracy only employed \gls{ppg} signal. Consequently, besides the physiological data, other parameters directly affect accuracy, such as preprocessing methods and classification algorithms.  
Table~\ref{table:ML_algos_table} 
shows the wide range of accuracy, 
between 72\% \cite{classification_of_human} and 90\% \cite{evaluation_of_learning}, for attention detection. 
Stress can be detected more precisely with accuracy ranging between 62\% \cite{personal_stress_level_clustering} and 95\% \cite{seongsil2021}. 
However, the used measurement devices for detecting attention, which often feature \gls{eeg} sensors, can be uncomfortable or intrusive for the users \cite{nonintrusive_vital_sign}. All of the researched studies use an ML algorithm for classification, and many compare different ones against each other. For filtering the raw data, on the other hand, a mixture of manual noise extraction, post-processing, and automated algorithms using accelerometers have been used ~\cite{evaluation_of_learning, detection_and_removal}. 

The practical use cases of the stress and attention detection methods mentioned before span 
from specialized workplaces like airplanes and law enforcement over athletes, or search and rescue missions to schools, where the attention level of 
students is often crucial   ~\cite{ photoplethysmographic_waveform_versus, dell2022machine,learning_immersion_assessment}. Stress detection on consumer devices can be used to avoid side effects of stress that can be detrimental to health, such as poor sleep~
\cite{personalized_stress_monitoring}. 

Determining with certainty which sensors or algorithms achieve the best results is challenging, as factors such as the noise level of the data, preprocessing methods, and algorithm parameters influence the outcomes. However, on average, ensemble algorithms, particularly XGBoost, consistently outperform other methods in stress detection. Their strength lies in combining multiple weak learners into a robust model. This also leads to reduced overfitting and enhanced generalization. In addition, XGBoost is less sensitive to noise compared to \gls{svm}, because \gls{svm} tries to find the optimal hyperplane that separates the data into different classes, but noise affects the position and the margin of the hyperplane.
As a result, since \gls{ppg} data generally includes more motion artifacts compared to \gls{eeg} signals, \gls{svm} demonstrates more promising results for attention detection than stress detection.  
Many new studies offer different options, with up and downsides to each of them, and although ML is always used, no single ML algorithm has been dominant in the literature. 
\\The experiment methodology varies in their setups, with participants' pool sizes ranging from 5 to 95. Especially, uncontrolled data measured in out-of-lab environments have high uncertainty and variation in quality as seen in Chapters \ref{stress_detection} and \ref{attention_detection}. 
\\
It is worth mentioning that certain studies employ \gls{dnn}s as their predictive models \cite{Abdo2021,human_attention_recognition,classification_of_human}. However, due to the requirement for substantial amounts of labeled data for training and longer training times in comparison with classical \gls{ml} models, \gls{dnn}s are harder to employ in wearable healthcare domain \cite{choi2020introduction,mukhamediev2021classical}. Data privacy concerns and high costs of data collection and storage are the reasons that make accessing large datasets more difficult \cite{slobogean2015bigger,aminifar2021privacy,aminifar2021diversity,aminifar2021scalable}.

In conclusion, the choice of suitable wearables highly relies on the specific application. For analyzing brain-related activities like attention level detection or cognitive load assessment, head-worn devices with various sensors (especially \gls{eeg}) are preferable. On the other hand, for stress detection, wrist-worn and finger-worn devices with diverse sensors (especially \gls{ppg}) are preferable. It's important to note that combining different physiological parameters is often helpful for more accurate attention and stress detection. Finally, there is not a specific prediction model which can consistently yield the best results.

\newacronym{d-t}{DT}{Decision Tree}
\newacronym{ert}{ERT}{Extremely Randomized Trees}
\newacronym{gbdt}{GBDT}{Gradient Boosted Decision Trees}

\section{Conclusion}
\label{sec_conclusion}

In this paper we studied various biomedical devices, including headbands, wristbands, smart rings, and chest bands, 
equipped with diverse biosensors 
such as \gls{ppg}, \gls{eeg}, \gls{eda}, and skin temperature sensors. In our investigation, we delved into the detailed analysis of each device, 
clarifying their respective capabilities and limitations. We 
presented the applications of these devices and sensors, particularly emphasizing their roles in physical and psychological assessments. That is, in early warning score systems (heart rate, blood pressure and respiratory rate monitoring) and 
the crucial domains of stress and attention detection, wherein we meticulously outlined the prevalent sensors frequently employed in the existing literature for these purposes. Stress and attention detection hold critical significance in various high-risk workplaces. 
Complex systems are necessary for accurate measurements, 
however, for enhanced comfort, despite potential limitations in data quality, head and wrist-worn devices present them as prime candidates. 

Furthermore, we studied machine learning algorithms commonly utilized for stress and attention detection, 
presenting their respective 
performance and limitations.
Despite the strides made in this area, our research highlighted the persistent challenges in obtaining reliable results, especially when implementing stress and attention detection systems within an organizational setting. For example, 
environmental dynamics of a workplace introduce complexities not encountered within controlled laboratory settings. Factors such as employee movement, talking, and other occupational activities can introduce significant noise and hinder the accuracy of detection systems.


We content that a portion of future efforts should 
concentrate on refining denoising methodologies to clean datasets efficiently, thereby mitigating the impact of environmental noise. Moreover, exploring robust machine learning algorithms that can operate effectively despite varying noise levels will be a critical avenue for further research, especially with the growing accessibility of new consumer-oriented devices that facilitate more extensive and inclusive studies. 
By addressing these challenges, we can pave the way for more robust and reliable stress and attention detection systems tailored for practical implementation within organizational contexts. A reliable and robust detection system is only the first step in improving the health of employees in the workplace. Without such reliable systems, it would be rather impractical for psychologist and sociologist to conduct (more) objective studies about parameters such as stress and devise solution that can use biological data to improve the health of workers and the working environment.



\section*{Acknowledgement}
This work has been partially funded by the Vienna Science and Technology Fund (WWTF) [10.47379/ICT20034].


\bibliographystyle{ACM-Reference-Format}
\bibliography{references.bib}


\end{document}